\documentclass[journal]{IEEEtran}

\usepackage{stfloats} % 强制插入表格在当前位置

\usepackage{cite}
\usepackage[cmex10]{amsmath}
\usepackage{amssymb}
\usepackage{cases}
\usepackage{graphicx}
\usepackage{mathtools}
\usepackage{multirow}
\usepackage{booktabs}
\usepackage{color}
\usepackage{tikz} 
\usetikzlibrary{arrows.meta, shapes.geometric, positioning}
\usepackage{algorithm}
\usepackage{makecell}
\usepackage{algorithmic}

\usepackage{changepage}

\usepackage{array}

\usepackage{url}
\usepackage{float}
\usepackage{verbatim}
\usepackage{soul}
\usepackage{amsthm,amsmath,amssymb}
\usepackage{mathrsfs}

\usepackage{bm}

\newtheorem{remark}{Remark}

\newcommand{\red}[1]{\textcolor{red}{#1}}

\newcommand{\bd}[1]{\textbf{#1}}

\soulregister\ref7 % 针对\ref命令
\soulregister\cite7 % 针对\ref命令

\renewcommand{\baselinestretch}{0.92}

\usepackage[normalem]{ulem}

\allowdisplaybreaks

\begin{document}

\title{Post-Earthquake Restoration of Electricity-Gas Distribution Systems with Damage Information Collection and Repair Vehicle Routing}

\author{Mingxuan Li, Wei Wei,~\IEEEmembership{Senior Member,~IEEE}, 
	Yin Xu,~\IEEEmembership{Senior Member,~IEEE}, Chengeng Zhang,~\IEEEmembership{Member, IEEE}, Shanshan Shi
	
	\thanks{This manuscript has been accepted for publication in \textit{CSEE Journal of Power and Energy Systems}.}
    \thanks{This work was supported by National Key Research and Development Program of China (Grant Number 2022YFB2405500) and the Science and Technology Project of State Grid Corporation of China (Grant Number 52094023001H). \textit{(Corresponding Author: Wei Wei.)}

    Mingxuan Li and Wei Wei are with Department of Electrical Engineering, Tsinghua University, 100084, Beijing, China. (e-mail: lmx20@mails.tsinghua.edu.cn, wei-wei04@mails.tsinghua.edu.cn)

    Yin Xu and Chengeng Zhang are with the School of Electrical Engineering, Beijing Jiaotong University, 100044,  Beijing, China. (e-mail: xuyin@bjtu.edu.cn, zhangchengeng@bjtu.edu.cn)

    Shanshan Shi is with State Grid Shanghai Municipal Electric Power Company, 200126, Shanghai, China. (e-mail: sss3397@163.com)
    }
}

\maketitle

\begin{abstract}
Extreme events such as earthquakes pose significant threats to integrated electricity-gas distribution systems (IEGDS) by causing widespread damage. Existing restoration approaches typically assume full awareness of damage, which may not be true if monitoring and communication infrastructures are impaired. In such circumstances, field inspection is necessary. This paper presents a novel adaptive restoration framework for IEGDS, considering dynamic damage assessment and repair. The restoration problem is formulated as a partially observable Markov decision process (POMDP), capturing the gradually revealed contingency and the evolving impact of field crew actions. To address the computational challenges of POMDPs in real-time applications, an advanced belief tree search (BTS) algorithm is introduced. This algorithm enables crew members to continuously update their actions based on evolving belief states, leveraging comprehensive simulations to evaluate potential future trajectories and identify optimal inspection and repair strategies. Based on the BTS algorithm, a unified real-time decision-making framework is developed for IEGDS restoration. Case studies on two distinct IEGDS systems demonstrate the effectiveness and scalability of the proposed method. The results indicate that the proposed approach achieves an outage cost comparable to the ideal solution, and reduces the total outage cost by more than 15\% compared to strategies based on stochastic programming and heuristic methods.
\end{abstract}


\section{Introduction}

The increasing frequency and intensity of extreme events have posed serious challenges to the reliable operation of critical energy infrastructures, highlighting the urgent need for enhanced system resilience \cite{resilience_importance_1}. In the context of the global transition toward cleaner energy sources, gas-fired generation units are playing an increasingly pivotal role in urban energy supply due to their flexibility and relatively low carbon emissions \cite{IEGDS_importance_1}. Consequently, the interdependence between electricity and natural gas systems has become more pronounced, underscoring the importance of improving the resilience of integrated electricity-gas distribution systems (IEGDS) \cite{IEGDS_resilience_metric_1, IEGDS_resilience_metric_2, IEGDS_resilience_market_2}.

Among various types of extreme events, earthquakes pose a particularly severe threat to IEGDS, as they can simultaneously cause widespread failures in both electric power infrastructure -- such as substations and transmission lines \cite{Earthquake_PS_Multishock} -- and in natural gas pipelines \cite{Earthquake_IEGDS_Impact}. Pipeline damage can lead to substantial gas supply shortages, which in turn compromise the operation of gas-fired power units and further degrade the performance of the power distribution network. This interlinked vulnerability reveals that enhancing the resilience of IEGDS involves more complex challenges \cite{IEGDS_restore_nonanticip_Zhang, IEGDS_restore_multi_task_Wang}. Therefore, it is imperative to develop methodologies for improving the resilience of coupled electricity-gas systems under seismic hazards.

The restoration of an integrated energy system has attracted increasing attention in recent years, with various works focusing on physical and operational aspects such as system reconfiguration, optimal crew dispatch, and considerations for information privacy during the recovery process. In \cite{IEGDS_restore_Shen}, a dynamic reconfiguration and operation approach is introduced for post-earthquake  IEGDS restoration. A particle swarm optimization algorithm is utilized to optimize system topology adjustments and operational strategies. In \cite{IEGDS_repair_Lin}, a joint repair scheduling model is introduced for power and gas systems, optimizing repair sequences to minimize load shedding costs and repair time. In \cite{IEGWDS_repair_Tajik}, a mixed-integer programming model is proposed to optimize the restoration of interdependent power, water, and gas networks by incorporating dynamic repair tasks to enhance resilience. In \cite{IEGDTS_repair_Wei, IEGDS_restore_ADMM_Li, IEGDS_restore_ADMM_Jing}, the restoration of power and gas networks is coordinated while preserving the privacy of the operational data of each system. Alternating Direction Method of Multipliers (ADMM) algorithms are employed for distributed optimization, enhancing recovery efficiency without requiring full information sharing between subsystems. \cite{IEGDS_restore_skeleten_1_Sang} proposes a restoration framework for interdependent gas and power networks, integrating skeleton-network reconfiguration and restoration sequence optimization into a unified MILP model. \cite{IEGDS_restore_MISOCP_Wang} considers the post-disaster scheduling of various mobile resources, including repair crews, fuel tankers, and mobile distributed generators. Dynamic traffic flow during restoration is considered in \cite{IEGDS_traffic_Jiang}.

While existing studies have provided valuable frameworks for post-disaster restoration, they typically assume complete and accurate knowledge of system damage before the repair process starts. However, this assumption may not hold in more severe events, such as an earthquake. Despite the growing intelligence of urban energy systems enabled by smart metering infrastructure, full awareness of damage at the beginning of restoration remains a big challenge. Practical limitations such as incomplete sensor coverage, communication failures, and physical destruction of sensing devices often hinder automated fault detection systems \cite{Modernizing_Chenchen, Restoration_Communication_Drone_Zhang, DS_assess_Wangzhaoyu_2025}. Consequently, field inspections are still indispensable for identifying undetectable faults. Given the poor road conditions after disasters \cite{post-disaster_debris_Celik} and the high labor intensity of inspection tasks \cite{line_inspection_Yang}, these field inspections are typically time-consuming and delay the overall recovery process. When damage assessment and repair are treated as sequential and decoupled tasks, restoration efforts may suffer from considerable inefficiencies.

To address the dynamic and uncertain nature of post-disaster environments, recent research has increasingly focused on integrating damage assessment with ongoing restoration efforts through rolling optimization strategies. These strategies enable iterative updates to repair schedules as new information emerges. For instance, \cite{DS_assess_restore_Bian} applies rolling re-optimization to adjust restoration plans for power networks based on newly discovered faults, leveraging technologies such as unmanned aerial vehicles. Similarly, \cite{DS_assess_restore_jalilian} develops a mixed-integer linear programming model that jointly optimizes feeder patrol, damage assessment, switching operations, and repair sequencing. However, a critical limitation of these efforts is their focus on power distribution systems, with limited attention paid to gas networks.

Gas systems present unique challenges that distinguish them from power systems. In particular, fault detection in buried gas pipelines generally requires on-site inspection when reliance on remote sensing technologies is limited after a disaster \cite{Pipeline_Detection_1}. While unmanned aerial vehicles can greatly accelerate damage assessment in overhead power lines \cite{UAV_inspection_Zhou}, similar advantages are not readily transferable to underground gas infrastructure, making inspection time-consuming.

Despite the growing attention to integrated energy systems, there is a noticeable gap in the literature on real-time restoration of IEGDS under conditions of partial observability. Effective strategies are required to guide field crews in performing both inspection and repair tasks in an integrated manner. This gap motivates the present study, which aims to develop a unified planning framework for IEGDS restoration that accounts for incomplete damage information and dynamically evolving system states. We model this problem using the partially observable Markov decision process (POMDP) framework to support real-time decision-making in the presence of uncertainty.

The key contributions of this study are as follows:

(1) \textbf{A novel POMDP-based model for joint damage assessment and restoration planning.}
We develop a framework that formulates the post-disaster restoration of IEGDS under partial observability as a POMDP. This formulation explicitly accounts for the necessity of field crews to simultaneously assess uncertain damage and execute repair operations. A belief state is applied to probabilistically estimate faults in uninspected areas, providing a robust foundation for adaptive and informed decision-making throughout the restoration process.

(2) \textbf{An advanced BTS algorithm for real-time gas crew dispatch.}
To address the computational challenges of POMDPs in real-time scenarios, we introduce a belief tree search (BTS) algorithm. This method allows crew members to continuously update their actions based on the evolving belief of the potential system states. By employing comprehensive simulations across various underlying scenarios and evaluating candidate actions, the BTS approach meticulously assesses potential future trajectories and computes corresponding Q-values, enabling the identification of optimal inspection and repair actions with high precision.

(3) \textbf{A unified real-time decision-making framework for integrated crew scheduling.}
This study presents a comprehensive real-time decision-making framework specifically designed for integrated crew scheduling. The framework seamlessly integrates the proposed BTS algorithm for gas crew scheduling with a rolling optimization approach tailored for power crew dispatch. This framework adapts to event-triggered decision updates for both power and gas crews, ensuring responsive and adaptive operations in response to real-time observations and dynamic system variations.

The remainder of this paper is structured as follows. Section \ref{section: prob_form} elaborates on the problem formulation for joint damage assessment and restoration in IEGDS. Section \ref{section: methodology} details the proposed BTS algorithm for gas crew scheduling and the rolling optimization strategy for power crew scheduling. Section \ref{section: case_study} presents comprehensive case studies based on two distinct IEGDS systems to validate the proposed methodologies. Finally, Section \ref{section: conclusion} concludes the paper and outlines promising directions for future research.

\section{Problem Formulation} \label{section: prob_form}

\subsection{Problem Description and Assumptions}
\label{subsec:problem_description}

Earthquakes may severely disrupt IEGDS, causing widespread outages. When the substation fails, the distribution network loses its connection to the main grid and needs to rely on distributed gas-fired units for emergency power supply. However, these generators may be inoperable due to damage in underground gas pipelines. Although advanced metering infrastructure can detect some faults, others may remain undetected due to failures in post-disaster monitoring and communication systems, necessitating on-site inspections. Under such partial observability, the system operator must develop an optimal inspection and repair strategy to maximize load restoration. This requires balancing the discovery of unknown faults with the repair of known ones -- particularly for gas crews responsible for both tasks. The following assumptions are made:

% In the event of an earthquake, the gas distribution network employs automated systems to ensure safety and facilitate restoration. Specifically, automatic shut-off valves are activated to isolate potentially damaged pipeline segments, preventing gas leaks and enabling unaffected areas to maintain or quickly resume supply. Pipelines with confirmed integrity, identified via real-time monitoring systems such as SCADA, are prioritized for immediate restoration. Conversely, pipelines with unknown status remain isolated until on-site inspection crews verify their safety and functionality, after which they are re-integrated into the operational network.

\begin{enumerate}
	\item \textbf{Partial observability:} Earthquakes damage both power and gas networks, impairing physical infrastructure and monitoring systems. Only partial damage can be assessed remotely; the rest requires on-site inspection.
	
	\item \textbf{Damage assessment and repair:}
	In power distribution networks, faults predominantly occur in lines and poles. To quickly identify unknown issues, unmanned aerial vehicles (UAVs) are deployed for rapid fault detection and reporting. Power crews then repair the confirmed faults with high priority \cite{DS_assess_restore_Bian}. In contrast, gas network faults primarily affect underground pipelines, where direct measurements are limited. As a result, gas crews must both inspect potentially damaged segments and repair confirmed faults. Pipeline failures often lead to methane leakage, which manifests as elevated methane concentrations and abnormal temperature profiles in the surrounding soil. To assess such conditions, gas crews use specialized equipment--such as methane detectors and infrared thermal imaging devices--for comprehensive soil inspection during field visits.

	% Unknown pipeline segments are assumed to be isolated by automatic valves and remain unavailable until inspected and verified by crews.
	
	\item \textbf{Steady-state assumption:} The restoration strategy is evaluated under steady-state conditions. Specifically, a damaged gas pipeline is assumed to fully interrupt gas flow until it is repaired, thereby cutting off the gas supply to downstream gas-fired generators. This assumption is reasonable for the urban medium- and low-pressure distribution networks considered in this study, where line-pack effects are limited: gas leakage from damaged pipelines is rapid and segments are promptly isolated by valves, while the residual volume in intact pipelines is insufficient to sustain generator operation over the 30-minute discretization adopted in the model.
\end{enumerate}

\begin{figure}[!t]
	\centering
	\includegraphics[width=0.8\linewidth]{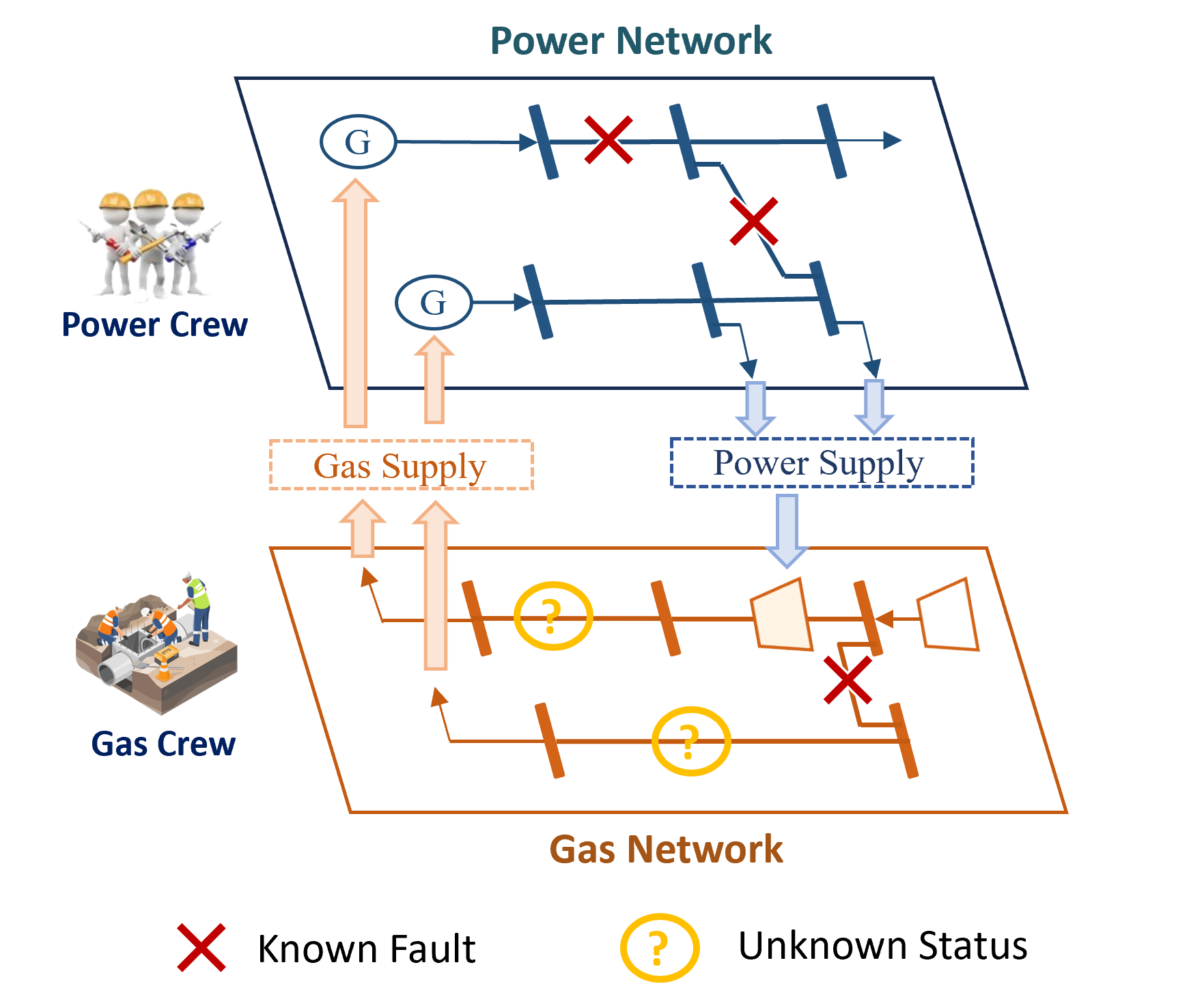}
	\caption{Illustration of IEGDS restoration under partial observability}
	\label{fig:IEGDS_restoration}
	% \vspace{-5pt}
\end{figure}

Under these assumptions, restoring the IEGDS becomes a decision-making problem under incomplete information. As illustrated in Fig. \ref{fig:IEGDS_restoration}, gas crews should decide between inspecting unknown pipelines or repairing known damaged ones, while power crews focus on restoring power system damages. The trapezoid in Fig.~1 denotes a possible electric-driven compressor, which may exist in some medium-pressure networks \cite{Compressor_1_OptiNet}. In cases without compressors, the model naturally reduces to a one-way gas-to-electric coupling through gas-fired units. After an earthquake, power crews and gas crews collaborate to restore the overall energy supply, ensuring that gas is delivered to gas-fired generators and end users, while electricity is supplied to power consumers. The following sections formulate the IEGDS restoration problem under incomplete information as a POMDP.

\subsection{Mathematical model of IEGDS operation}
We first elaborate on the mathematical model of IEGDS operation under perfect information and point out how the schedule of repair crews affects the restoration process. Then, we transform the restoration problem into a POMDP, as not all the damages are initially acquired.

\subsubsection{Power Distribution Network Model}

The power flow in the distribution network is described using the linearized DistFlow model \cite{DistFlow}, as detailed below:
\begin{subequations} \label{eq:DN_power}
	\begin{gather}
	0 \leq p_{i, t}^{DG} \leq \overline{{P}_{i}^{DG}}, \  0 \leq q_{i, t}^{DG} \leq \overline{{Q}_{i}^{DG}}, \forall i, \forall t \label{eq:DN_8}\\
	0 \leq p_{i, t}^{D} \leq P_{i,t}^D, \  0 \leq q_{i, t}^{D} \leq Q_{i,t}^D, \forall i, \forall t \label{eq:DN_load} \\ 
	p_{i, t}^{DG} + \sum_{\mathclap{\forall (k, i) \in \Omega^L}} p_{ki, t}^{\ell} = p_{i, t}^{D}+p_{i, t}^{comp} +\sum_{\mathclap{\forall (i, j) \in \Omega^L}} p_{ij, t}^{\ell}, \forall i, \forall  t \label{eq:DN_3}\\
	q_{i, t}^{DG} + \sum_{\mathclap{\forall (k, i) \in \Omega^L}} q_{ki, t}^{\ell} = q_{i, t}^{D}+\sum_{\mathclap{\forall (i, j) \in \Omega^L}} q_{ij, t}^{\ell}, \forall i, \forall t \label{eq:DN_4}\\
	\begin{aligned}
	\left| V_{j, t}-V_{i, t} + {R_{\ell} p_{ij, t}^{\ell}+X_{\ell} q_{ij, t}^{\ell}}/{V_0} \right|  \leq  M \left(1 -z_{\ell, t}^L\right), \quad \\ \forall \ell, \forall t  
	\end{aligned} \label{eq:DN_5} \\
	V_i^l \leq V_{i, t} \leq V_i^u, \forall i, t \label{eq:DN_7}\\
	-z_{\ell, t}^L \overline{P_{\ell}^L}  \leq p_{\ell, t}^{L} \leq z_{\ell, t}^L \overline{P_{\ell}^L}, \forall \ell, \forall t \label{eq:DN_10}\\
	-z_{\ell, t}^L \overline{Q_{\ell}^L} \leq q_{\ell, t}^{L} \leq z_{\ell, t}^L \overline{Q_{\ell}^L}, \forall \ell, \forall  t \label{eq:DN_11}
	\end{gather}
\end{subequations}
where $p_{i,t}^{DG}/q_{i,t}^{DG}$ denotes the active/reactive power output of distributed generators (DGs) at bus $i$ at time $t$, bounded above by $\overline{P_{i}^{DG}}$ and $\overline{Q_{i}^{DG}}$. $p_{i,t}^D$ and $q_{i,t}^D$ represent the supplied power loads at bus $i$, corresponding to the load demands $P_{i,t}^D$ and $Q_{i,t}^D$, respectively. Compressor consumption at node $i$ is modeled as $p_{i, t}^{comp}$. Power flow of line $\ell$, connecting buses $i$ and $j$, is represented by $p_{ij,t}^{\ell}$ and $q_{ij,t}^{\ell}$. Line parameters $R_\ell$ and $X_\ell$ denote resistance and reactance, while $V_{i,t}$ is the bus voltage, constrained between $V_i^l$ and $V_i^u$. Binary variable $z_{\ell,t}^L$ indicates whether line $\ell$ is operational at time $t$, and $M$ is a sufficiently large constant used in constraint relaxation.

Constraint (\ref{eq:DN_8}) regulates DG power output, while (\ref{eq:DN_load}) ensures the supplied load remains within the demand. Power balance at each bus is modeled via (\ref{eq:DN_3}) and (\ref{eq:DN_4}). Voltage drop is governed by constraint (\ref{eq:DN_5}), and acceptable voltage limits are imposed in (\ref{eq:DN_7}). Constraints (\ref{eq:DN_10}) and (\ref{eq:DN_11}) restrict power flow on each line based on its operational status.

\subsubsection{Gas Network Model}

The operation of the gas system adheres to the following constraints:
\begin{subequations} \label{eq:GN}
	\begin{gather}
	W^{GW}_{i, \text{min}} \leq w^{\text{GW}}_{i, t} \leq W^{GW}_{i, \text{max}}, \quad \forall i \in GW, \forall t \label{eq:GN_1} \\
	0 \le w^{D}_{i,t} \le W^{D}_{i,t}, \forall i \in GN, \forall t  \label{eq:GN_2}\\
	\begin{aligned}
	w_{i, t}^{\text{GW}} - w^{D}_{i,t} - w^{DG}_{i,t} = \sum_{m \in \Omega(i)} f_{m,t} - \sum_{k \in \Pi(i)} f_{k,t}, \quad \\ \forall i \in GN, \forall t 
	\end{aligned}\label{eq:GN_3}\\
	\pi_{i,t} \leq \pi_{j,t} \leq \lambda_m \pi_{i,t}, \quad \forall m \in \Gamma_{\text{act}}, \forall t \label{eq:GN_4}\\
	f_{m,t}^2 - \phi_m (\pi_{i,t}^2 - \pi_{j,t}^2) = 0, \quad \forall m \in \Gamma_{\text{ina}}, \forall t \label{eq:GN_5}\\
	0 \leq f_{m,t} \leq F^{\text{max}}_{m}  (1 - z^W_{m,t}), \quad \forall m \in \Gamma_{\text{act}}\cup\Gamma_{\text{ina}}, \forall t \label{eq:GN_6}
	\end{gather}
\end{subequations}
where $w^{\text{GW}}_{i, t}$ indicates the gas input of well $i$ at time $t$, within bounds $W^{GW}_{i,\min}$ and $W^{GW}_{i,\max}$ for all gas well nodes $i \in GW$. The gas demand and supply at node $i$ are represented by $W^D_{i,t}$ and $w^D_{i,t}$, respectively. The gas consumed by DGs at node $i$ is denoted by $w^{DG}_{i,t}$. Sets $\Omega(i)$ and $\Pi(i)$ define the pipelines connected upstream and downstream of node $i$. $f_{m,t}$ is the gas flow through pipe $m$ at time $t$, and $\pi_{i,t}$ is the pressure at node $i$. Compression behavior on active pipelines $\Gamma_{act}$ is represented using ratio $\lambda_m$, and inactive pipelines (set $\Gamma_{ina}$) follow the Weymouth-type relation, parameterized by $\phi_m$. Binary variable $z^W_{m,t}$ tracks the operational state of pipeline $m$, and $F^{\max}_m$ defines its flow capacity.

Constraints (\ref{eq:GN_1}) and (\ref{eq:GN_2}) limit well output and gas load supply. Nodal gas balance is ensured by (\ref{eq:GN_3}). Constraints (\ref{eq:GN_4})--(\ref{eq:GN_5}) represent the physical pressure-flow relationships for both active and inactive pipelines, while (\ref{eq:GN_6}) imposes flow limits.

\subsubsection{Coupling Constraints}

The interdependence between the electricity and gas networks is modeled through the fuel usage of gas-fired DGs and the electrical load of gas compressors:
\begin{subequations} \label{eq:coupling}
	\begin{gather}
	w^{DG}_{m,t} = \beta_i P^{DG}_{i, t} + \gamma_i, \quad \forall i \in \Gamma_{DG}, \forall t \label{eq:gas_power_conv}  \\ 
	P^{comp}_{i, t} = \zeta_m f_{m, t}, \quad \forall m \in \Gamma_{\text{act}}, \forall t 
	\end{gather}
\end{subequations}
where parameters $\beta_i$ and $\gamma_i$ capture the gas consumption characteristics of each gas-fired DG, while $\zeta_m$ represents the electricity usage rate of the compressor installed on pipeline $m$. The above equations establish a bi-directional dependency between the two systems.

Collectively, constraints (\ref{eq:DN_power})–(\ref{eq:coupling}) characterize the feasible operating set of the IEGDS. Notably, constraint (\ref{eq:GN_5}) introduces nonlinearity due to its quadratic nature, but it can be approximated via piecewise linearization techniques as described in \cite{PWL_gas_model}, thus transforming the entire model into a mixed-integer linear program.

\subsubsection{IEGDS Restoration Problem}

Considering the post-disaster scenario, the objective is to minimize the overall weighted loss of electrical and gas loads:
\begin{equation}  \label{eq:IEGDS_operation}
\begin{aligned}
\min ~~ & \sum_t \left(C_t^P + C_t^W \right) \Delta t  \\
\mbox{s.t.} ~~ & \eqref{eq:DN_power} - \eqref{eq:coupling}
\end{aligned}
\end{equation}
where $C_t^P=\sum_{i \in \mathcal{L}_P} c_{i}^P \left(P_{i,t}^D - p_{i,t}^D\right)$ and $C_t^W =  \sum_{j \in \mathcal{L}_G} c_{j}^W  \left( W_{j,t}^D - w_{j,t}^D\right)$ represent the single-period load shedding costs for power and gas loads, respectively. $\mathcal{L}_P$ represents the set of buses in the power network, and $\mathcal{L}_G$ denotes the set of gas nodes. $c_{i}^P/c_{j}^W$ is the unit shedding cost for power$/$gas load at node $i/j$. 

The restoration problem \eqref{eq:IEGDS_operation} seeks to restore critical services while accounting for the real-time availability of power lines and gas pipelines, whose operational states, $z_{\ell, t}^L$ and $z_{m,t}^W$, evolve as repairs progress. Importantly, the inspection and restoration of these components are closely linked to the scheduling of power and gas crews. However, not all failures are known in advance--many are only discovered during field inspections. As a result, the system state is only partially observable and unfolds dynamically over time.

This inherent uncertainty and the sequential nature of information acquisition make it challenging to model the restoration process as a deterministic optimization problem. To address this, we introduce a POMDP framework, which supports sequential decision-making under uncertainty and partial observability. It naturally captures the interdependence between component restoration, system state evolution, and crew deployment.

\subsection{POMDP Formulation for Fault Restoration}

To handle the sequential decision-making under incomplete information, the restoration of IEGDS is formulated as a POMDP, defined by the tuple $\langle \mathcal{S}, \mathcal{A}, \mathcal{O}, T, O, C, b_0 \rangle$, where:

\textbf{1) State Space $\mathcal{S}$}: 
Each system state $s_t \in \mathcal{S}$ represents the complete operational and logistical status at time $t$. It comprises the binary operational status of infrastructure components—specifically, the condition of power lines \( z^L_{\ell,t} \) and gas pipelines \( z^W_{m,t} \), as well as the real-time status of all repair crews. The state of a repair crew \( c \) is represented as \( s^C_{c,t} = \{\alpha^C_{c,t}, u^C_{c,t}, \tau^C_{c,t} \} \), where \( \alpha^C_{c,t} \) denotes the current component that crew $c$ is working on (e.g., a damaged line or pipeline), \( u^C_{c,t} \in \{0,1\} \) indicates whether the crew is actively performing repairs (\( u=1 \)) or en route to the target (\( u=0 \)), and \( \tau^C_{c,t} \) specifies the remaining travel time to reach the target. Repair crews are categorized into two types: power crews and gas crews, with the crew type denoted by \( C \in \{PC, GC\} \).

\textbf{2) Action Space $\mathcal{A}$}: 
An action $a_t=\{\beta^C_{c,t}, \forall c\}$ specifies the subsequent target assignments for each crew $c$. Specifically, once a crew completes its current task, it is dispatched to a new target component $\beta^C_{c,t} \in \mathcal{Z}$, where $\mathcal{Z}$ denotes the set of all unknown or faulty components.

\textbf{3) Observation Space $\mathcal{O}$}:  
The observation $o$ comprises observed fault information, gas shortage data from gas-fired units, and end-user outage notifications. Furthermore, pipeline statuses verified through gas crew inspections are dynamically incorporated into $o$. Based on these observations, the conditions of components $z^L_{\ell,t}$ and $z^W_{m,t}$ can be partially inferred. However, the exact states of components that have not yet been inspected remain uncertain.

\textbf{4) State Transition Function $T(s'|s, a)$}:  
System transitions from state $s_t$ to $s_{t+1}$ are driven by component recovery dynamics and crew movement. The recovery status of each component is updated based on whether a repair crew is actively assigned to it and the duration of the ongoing repair. Crew state transitions involve either continuing their travel or initiating repair tasks, depending on their new assignments and the corresponding travel times.

The evolution of each crew's status is described by:
\begin{align}
\alpha_{c,t+1}^{\mathcal{C}} &= 
\begin{cases}
\emptyset, & \text{if } u_{c,t}^{\mathcal{C}}=0 \land \tau_{c, t+1}^{\mathcal{C}} > 0 \\ 
\beta_{c,t}^{\mathcal{C}}, & \text{if } u_{c,t}^{\mathcal{C}}=0 \land \tau_{c, t+1}^{\mathcal{C}} = 0 \\ 
\alpha_{c,t}^{\mathcal{C}}, & \text{if } u_{c,t}^{\mathcal{C}}=1 
\end{cases}  \label{eq:crew_trans_1} \\[5pt]
u_{c,t+1}^{\mathcal{C}} &= \mathbb{I} \Biggl(
  \begin{aligned}
    &\Bigl[ \alpha_{c,t+1}^{\mathcal{C}} = \beta_{c,t}^{\mathcal{C}} \lor u_{c,t}=1 \Bigr] \\ 
    &\land \Bigl[ r_{(\alpha_{c,t+1}^{\mathcal{C}}), t+1} < \mathcal{R}_{(\alpha_{c,t+1}^{\mathcal{C}}),t} \Bigr]
  \end{aligned}
\Biggr)  \label{eq:crew_trans_2} \\[5pt]
\tau_{c, t+1}^{\mathcal{C}} &= 
\begin{cases} 
\max\{0, \tau_{c, t}^{\mathcal{C}} - 1 \}, & \text{if } u_{c,t}^{\mathcal{C}}=0 \\ 
\mathcal{T}^{\mathcal{C}}(\alpha_{c,t}^{\mathcal{C}}, \beta_{c,t}^{\mathcal{C}}), & \text{if } u_{c,t}^{\mathcal{C}}=1 
\end{cases}  \label{eq:crew_trans_3}
\end{align}
where $\beta_{c,t}^{\mathcal{C}}$ is the new target assigned by the action, $\mathcal{T}^{\mathcal{C}}(\cdot,\cdot)$ is the crew-specific travel time between components, $r_{(x),t}$ is the accumulated inspection/repair time on component $x$ up to time $t$,  
 $\mathcal{R}_{(x),t}$ is the total inspection/repair time required to restore component $x$, $\mathbb{I}(\cdot)$ is the indicator function that returns 1 if the condition is satisfied, and 0 otherwise. Eq.~\eqref{eq:crew_trans_1} updates the component that the crew is currently working on. Eq.~\eqref{eq:crew_trans_2} updates the working status of the crew based on whether it has arrived at the assigned target and the progress made toward completing the required repair time. Eq.~\eqref{eq:crew_trans_3} updates the remaining travel time, either decrementing it if the crew is still en route or resetting it based on the distance to a newly assigned target. Together, these equations describe the evolution of crew states, capturing the impacts of task assignments and logistical constraints over time.

Especially, once the cumulative repair time $r_{(x),t}$ reaches the threshold ${\mathcal{R}}_{(x),t}$, the component is considered restored and its status variable $z_{x,t}$ transitions from 0 to 1.  The crew then proceeds to its next assigned target $\beta_{c,t}^C$. 

\textbf{5) Observation Function $O(o|s', a)$}: The observation function defines the probability of receiving observation $o$ after taking action $a$ and transitioning to state $s'$. In the context of IEGDS restoration, this corresponds to updating the status of a gas pipeline based on the inspection results provided by gas crews. Specifically, if a crew inspects a pipeline and confirms it is operational, the observation accurately reflects this updated status. Conversely, if a pipeline is not inspected by a crew, its status remains uncertain, reflecting the inherent probabilistic nature of uninspected components.

\textbf{6) Cost Function $C(s, a)$}: The cost function at each time step $t$ is $C(s_t, a_t) = \left(C_t^P + C_t^W \right) \Delta t$, corresponding to the single-step loss in objective~\eqref{eq:IEGDS_operation}.

\textbf{7) Initial Belief $b_0$}: This function encapsulates the probabilistic estimation of the initial system state, based on the currently available information and observation.

Based on the above information, the decision-maker aims to identify a policy $\pi: \mathcal{B} \rightarrow \mathcal{A}$ that minimizes the expected cumulative loss:
\begin{equation}
\label{eq:Restoration-multiperiod}
\begin{aligned}
V^\pi(b) = & \min \mathbb{E} \left[ \sum_{t=1}^T  C(s_t, a_t) \mid b_0 = b, \pi \right] 
\end{aligned}
\end{equation}
with belief updates governed by the standard Bayesian filter:
\begin{equation}
b'(s') = \frac{O(o|s', a) \sum_{s \in \mathcal{S}} T(s'|s, a) b(s)}{P(o|a, b)}
\end{equation}

Given the scarcity of historical data on extreme disasters and the high-dimensional complexity of crew routing decisions, training a generic model offline with limited data is challenging. In the following section, we present an online decision-making algorithm without offline training that is both efficient and practical for implementation. It continually updates each crew's immediate target, taking into account the impact of current decisions on future trajectories and objectives.

\section{Methodology} \label{section: methodology}

\begin{figure*}[!hbtp]
    \centering
    \includegraphics[width=0.9\linewidth]{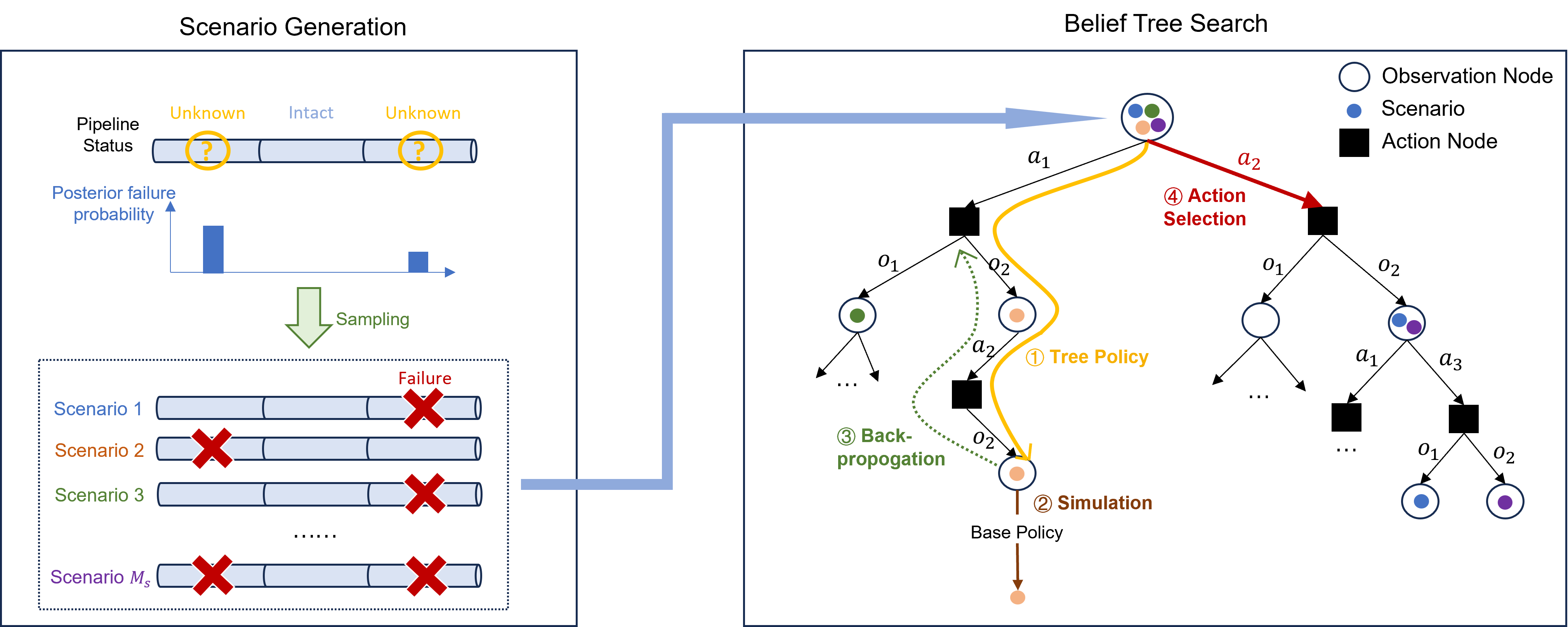}
    \caption{Belief tree search for gas crew scheduling}
    \label{fig:AlgTree}
    % \vspace{-5pt}
\end{figure*}

This section presents a decision-making framework for IEGDS restoration under partial observability. A key challenge lies in identifying the optimal targets for gas crews when the fault status of the gas network is only partially known. To address this, the BTS algorithm is proposed, as illustrated in Fig.~\ref{fig:AlgTree}. Based on estimated fault probabilities, a set of representative scenarios is generated to capture the current belief over the system state. A tree search is then conducted across these scenarios to simulate restoration trajectories under different actions. The action that minimizes the expected outage cost under the current belief is selected for execution.

The framework is elaborated in the subsequent sections. Section~\ref{sec:method_post_prob} describes the Bayesian inference and scenario generation process used to estimate the fault probabilities of uninspected pipelines. Section~\ref{section: tree_search} details the core BTS algorithm for gas crew scheduling. After real-time gas crew decisions are made, the schedule of power crews is updated accordingly, as presented in Section~\ref{sec:alg_power_crew}. Additional details, including energy flow approximation and multi-crew coordination, are introduced in Sections~\ref{sec:flow_approx} and~\ref{sec:multi_gas_crew}, respectively. Finally, the overall decision-making framework is summarized in Section~\ref{section: overall_framework}.

% To handle this challenge, the probabilities of unknown pipeline status is first estimated using Bayesian inference, using side information to estimate pipeline fault proabbilityes more credibly and accurately. 

% For gas network restoration, the unknown status of pipelines is first estimated using Bayesian inference. Based on the estimated fault probabilities, the proposed BTS algorithm is applied for gas crew scheduling under partial information, where numerous simulations are performed to evaluate candidate actions across sampled scenarios. Meanwhile, power crew routing is managed through a rolling optimization approach based on the progress of gas network restoration.  To efficiently simulate the restoration process within the BTS algorithm, an approximate algebraic model is proposed to represent integrated energy flows, relieving the computational burden associated with solving optimization problems during simulation. Additionally, since the standard tree search algorithm is not well-suited for multi-agent coordination, a sequential decision-making approach is employed to handle the scheduling of multiple gas crews effectively.

\subsection{Probabilistic Estimation of Pipeline Faults} \label{sec:method_post_prob}

As shown in Fig.~\ref{fig:AlgTree}, the proposed BTS algorithm requires a set of generated scenarios as input, where each scenario represents a possible configuration of uninspected pipeline states. These scenarios are constructed based on fault probabilities estimated from available information.

Following an earthquake, the prior failure probability of a pipeline $m$ is estimated using a Poisson distribution model \cite{Pipe_Damage_Prob_1, Pipe_Damage_Prob_2}, given by:
\begin{equation} \label{eq:pipe_prior_prob}
    p_m^0 = 1 - e^{-0.00003 \times (\text{PGV})^{2.25} \times L_m}
\end{equation}
where PGV represents the peak ground velocity of the earthquake, and $L_m$ is the length of the pipeline $m$.

However, prior probabilities alone do not account for post-disaster observations such as user-reported outages or gas shortages at generators. To incorporate such information, Bayesian inference is applied to obtain posterior estimates. Let \( \mathbf{z} = [z_1, \dots, z_N] \) denote the binary fault vector for all pipelines, where \( z_m = 0 \) indicates that pipeline \( m \) has failed, and \( p_m^0 \) is its prior failure probability. The set \( \mathcal{O}(\mathbf{z}) \) represents the unserved nodes under configuration \( \mathbf{z} \), as determined by gas flow simulation. Then, given the observed set of unserved gas nodes \( \mathcal{U} \), the posterior probability that pipeline \( m \) is in a failed state is computed as
\begin{equation} \label{eq:pipe_post_prob}
P(z_m = 0 \mid \mathcal{U}) = 
\frac{
	\sum\limits_{\mathbf{z} \in \{0,1\}^N} 
	\mathbb{I}[z_m = 0] \cdot 
	\mathbb{I}[\mathcal{U} \subseteq \mathcal{O}(\mathbf{z})] \cdot 
	P(\mathbf{z})}{
	\sum\limits_{\mathbf{z} \in \{0,1\}^N} 
	\mathbb{I}[\mathcal{U} \subseteq \mathcal{O}(\mathbf{z})] \cdot 
	P(\mathbf{z})}
\end{equation}
where \( \mathbb{I}[\cdot] \) is the indicator function. The prior joint probability of pipeline states $P(\mathbf{z}) = \prod_{m=1}^{N} (p_m^0)^{1 - z_m} (1 - p_m^0)^{z_m}$. Eq.~\eqref{eq:pipe_post_prob} evaluates the posterior by marginalizing over all system configurations that are consistent with the observed unserved gas nodes \( \mathcal{U} \), conditioned on pipeline \( m \) being in a failed state.

Let $\phi_m = P(z_m = 0 \mid \mathcal{U})$ denote the posterior failure probability of pipeline~$m$. If its actual status is known (i.e., confirmed failed or operational), then $\phi_m$ is fixed to $1$ or $0$ accordingly. Based on $\{\phi_m\}_{m=1}^N$, we generate a belief set by sampling \( M_s \) scenarios, each corresponding to a possible realization of unknown pipeline states. These scenarios initialize the root node of the belief tree used in the tree search algorithm (see Section~\ref{section: tree_search}). In each simulation, the sampled states override unknown values, and the sampling frequency reflects the posterior likelihood, providing a probabilistically grounded basis for planning and decision-making under uncertainty.

\subsection{Belief Tree Search for Gas Crew Scheudling} \label{section: tree_search}

Based on the sampled scenarios of unknown pipelines states, the BTS algorithm is employed for gas crew scheduling under the current belief. The tree structure, illustrated in Fig.~\ref{fig:AlgTree}, consists of alternating \emph{action nodes} and \emph{observation nodes}. Each action node encodes a decision to dispatch the gas repair crew to inspect or repair a target pipeline. Inspection and repair are modeled jointly as a compound action, consistent with reality where crews immediately repair a detected fault. The child observation nodes of each action node represent the possible observed states of the target pipeline: \( o_1 \) (intact) or \( o_2 \) (faulty). Scenarios following the same action diverge into different branches based on their respective observations, which may imply different completion durations and thus load loss trajectories.

The search process follows a Monte Carlo tree search structure with domain-specific adaptations. The overall procedure is illustrated in Fig.~\ref{fig:AlgTree} and summarized in Algorithm~\ref{alg:gas_pomdp}, with key steps detailed below:

\begin{enumerate}
\item \textbf{Tree Policy} \textnormal{(Algorithm 1 lines 6–13)}:  
Starting from the root node, the search tree is recursively traversed via branch selection using an Upper Confidence Bound (UCB) rule:
\[
\hat{a} = \arg\min_a \left(Q(a) - c \sqrt{\frac{\log N_{\text{parent}}}{N(a)}}\right),
\]
where $\hat{a}$ is the action branch selected by the UCB rule, \( Q(a) \) is the average cumulative cost of action \( a \), and \( N(a) \) its visit count. If the selected action leads to an unvisited observation node, a new belief node is added to the tree. Progressive widening \cite{MCTS_survey} is used to limit the number of candidate actions, gradually expanding the search space based on node visit counts.

\item \textbf{Simulation} \textnormal{(Algorithm 1 line 14)}:  
Once the search reaches a leaf node or the maximum tree depth, the remaining horizon is simulated using a heuristic base policy \( \pi_{\text{base}} \). This policy selects the pipeline \( j^* \) maximizing load gain per unit time:
\[
j^* = \arg\max_j \left( \text{LoadGain}_j / \mathcal{R}_j \right).
\]

Using the base policy, the subsequent restoration sequence is simulated by estimating crew travel and repair progress over the remaining horizon. In practical applications, the travel-time and workload parameters used in the simulation can be obtained from real-time traffic and field data.

\item \textbf{Backpropagation} \textnormal{(Algorithm 1 lines 15–16)}:  
The cumulative cost from the simulation is propagated backward along the visited trajectory. Update action statistics by:
\[
Q(a) \leftarrow \frac{N(a) Q(a) + \sum_{t_a}^{T} C}{N(a) + 1}, \quad N(a) \leftarrow N(a) + 1,
\]
where \( t_a \) is the time at which action \( a \) is executed.

\item \textbf{Action Selection} \textnormal{(Algorithm 1 line 18)}:  
After all scenarios have been evaluated, the algorithm selects the root-level action with the lowest expected cost:
\[
a^* = \arg\min_{a \in \mathcal{A}_1} Q(a)
\]
where $\mathcal{A}_1$ denotes the set of candidate actions in the first layer of the decision tree. This action is executed in the real system, and the belief set is updated accordingly.

\end{enumerate}

\begin{remark}[Event-Based Cost Modeling]
Unlike conventional fixed-step decision problems (e.g., games or time-slot dispatch), our restoration actions are event-driven and may span a variable number of time steps. For example, inspecting a pipeline lasts only for the inspection period if it is intact, but extends to include both inspection and repair if it is damaged. Outage costs are accumulated at fixed intervals and aggregated over the entire duration $[t_a, t_a+\tau_a]$ of each action $a$. This event-based attribution avoids fragmenting the cost of a single action across multiple steps and establishes a clear one-to-one correspondence between each action and its cumulative impact in belief tree search.
\end{remark}

\begin{remark}[Simulation of Energy Flow]
The proposed BTS method requires extensive simulation of IEGDS energy flows to estimate cumulative load loss over multiple time steps across a large number of scenarios. Solving optimization problems for each simulation would reduce computational efficiency and hinder parallelization. To address this, we introduce an algebraic approximation algorithm for IEGDS flow estimation that is both fast and effective. Detailed descriptions are provided in Section~\ref{sec:flow_approx}.
\end{remark}

\begin{remark}[Integration with Multi-Agent Restoration]
While the belief tree models the decision-making process of a single gas crew, each simulation within the tree integrates concurrent actions from other agents, including power crews and additional gas crews. Power crew actions are assumed to be pre-determined through a rolling horizon optimization (see Section~\ref{sec:alg_power_crew}), while the scheduling of other gas crews follows predefined priority heuristics (see Section~\ref{sec:multi_gas_crew}). This hybrid coordination ensures that simulated trajectories reflect realistic multi-agent interactions, allowing for more accurate and actionable planning outcomes.
\end{remark}

\begin{algorithm}[htbp]
\caption{Belief Tree Search for Gas Crew Scheduling}
\label{alg:gas_pomdp}
\begin{algorithmic}[1]
\STATE Sample $M_s$ failure scenarios from posterior distribution;
\STATE Initialize root belief node using current system state and $M_s$ scenarios;
\FOR{each scenario $m = 1, \ldots, M_s$}
    \STATE Sample a realization of the system state based on scenario $m$;
    \STATE Start from root node;
    \WHILE{not reaching maximum depth}
        \STATE Select action $a$ using UCB from current belief node;
        \IF{action $a$ leads to unvisited observation}
            \STATE Create new observation node $o$ and corresponding belief node;
        \ELSE
            \STATE Transition to next node based on observation $o_m$;
        \ENDIF
    \ENDWHILE
    \STATE Simulate remaining horizon using base policy $\pi_{\text{base}}$;
    \STATE Compute cumulative cost $\sum_{\tau_a}^{T} C$;
    \STATE Backpropagate cost: update $Q(a)$ and $N(a)$ along visited nodes;
\ENDFOR
\STATE Select $a^* = \arg\min Q(a)$ as the best root action.
\end{algorithmic}
\end{algorithm}

\subsection{Rolling Optimization for Power Crew Scheduling} \label{sec:alg_power_crew}

To handle the interdependence between gas and power system recovery, we propose a rolling optimization framework for power crew scheduling. It dynamically adapts to newly detected failures and updated gas network status, enabling coordinated repair planning for system-level recovery.

When power crews need to update the schedule at time step $t$, a mixed-integer program optimizes crew routing and repair scheduling over the remaining horizon based on the latest gas generator availability while fixing all prior decisions. The objective is to minimize the total cost of load loss starting from the current time step $t$, i.e.,
\begin{equation} \label{eq:power_crew_rolling_opt}
\begin{aligned}
\min ~~ & \sum_{\tau=t}^{T} \sum_{i \in \mathcal{L}_P} c_i^P (P^D_{i, \tau} - p^D_{i, \tau}) \\
\mbox{s.t.} ~~ &  \eqref{eq:DN_power}, \eqref{eq:power_route_x} - \eqref{eq:historical_route}
\end{aligned}
\end{equation}
where the following constraints are considered:
\subsubsection{Power crew routing constraints}
\begin{subequations} \label{eq:power_route_x}
\begin{align} 
    \sum_{j \in \mathcal{F}} x_{0j}^k &= 1, \forall k \in \mathcal{K} \\
    \sum_{j \in \mathcal{F}} x_{ij}^k &\leq 1, \forall i \in \mathcal{F},\ \forall k \in \mathcal{K} \\
    \sum_{i \in \mathcal{F}} x_{ij}^k &\leq 1,  \forall j \in \mathcal{F},\ \forall k \in \mathcal{K} \\
    \sum_{j \in \mathcal{F}} x_{ij}^k &= \sum_{j' \in \mathcal{F}} x_{jj'}^k, \forall i \in \mathcal{F},\ \forall k \in \mathcal{K}
\end{align}
\end{subequations}
where \( x_{ij}^k \in \{0,1\} \) indicates whether crew \( k \) travels from component \( i \) to \( j \), and \( \mathcal{F} \) includes all faulty lines. Constraint (\ref{eq:power_route_x}) ensures that each repair crew departs from its depot, visits a subset of faulty lines, and follows a continuous route.

\subsubsection{Arrival and repair time constraints}
\begin{subequations}
\begin{align}
  |\tau_j^k - \tau_i^k - R_i - D_{ij}| \leq & (1 - x_{ij}^k) M,  ~~ \forall i, j \in \mathcal{F}  \label{eq:arrival_time} \\
  \sum_{t = 1}^{T} h_{i,t} = & 1 ~~~~~~~~~~~~~~~~ \  \forall i \in \mathcal{F} \label{eq:unique_repair_time} \\
  \sum_{t = 1}^{T} t \cdot h_{i,t} = \tau_i^k + R_i \cdot & \sum_{j \in \mathcal{F}} x_{ij}^k, ~~  \forall i \in \mathcal{F},\ \forall k \in \mathcal{K} \label{eq:repair_completion} \\
  z_{i,t}^L = \sum_{\tau=1}^{t} h_{i, \tau} &, ~~~~~~~ \forall i \in \mathcal{F} ~ \forall t \in [1, T] \label{eq:line_state_after_repair}
\end{align}
\end{subequations}
where \( \tau_i^k \) denotes the arrival time of crew \( k \) at component \( i \); \( R_i \) is the required repair duration for component \( i \); and \( D_{ij} \) represents the travel time between components \( i \) and \( j \); $h_{i,t}$ indicates whether the repair of component $i$ is completed at time $t$. Constraint \eqref{eq:arrival_time} ensures temporal consistency between visited components for each crew; \eqref{eq:unique_repair_time} ensures each fault is repaired exactly once; \eqref{eq:repair_completion} ties the declared restoration time to the actual crew arrival and repair duration; \eqref{eq:line_state_after_repair} ensures that a line becomes operational after it has been repaired.

\subsubsection{Gas-aware DG constraints}
\begin{equation}
   P^{DG}_{i, \tau} \leq (\hat{W}^{DG}_{m,\tau} - \gamma_i ) / \beta_i, \quad \forall i \in \Gamma_{DG}, \ \tau \in [t,T]  \label{eq:gas_power_conv_2}
\end{equation}
where $\hat{W}^{DG}_{m,\tau}$ denotes the estimated gas supply available to gas-fired generator \( m \), obtained from gas flow simulation results based on the belief tree search.

\subsubsection{Historical consistency constraints}
\begin{equation}  \label{eq:historical_route}
    x_{ij}^k = x_{ij}^{k,\text{prior}}, \quad 
    \tau_j^k = \tau_j^{k,\text{prior}}, \quad 
    \forall \tau_j^{k,\text{prior}} \leq t
\end{equation}
where \( x_{ij}^{k,\text{prior}} \) and \( \tau_j^{k,\text{prior}} \) represent the routing decisions and arrival times that were finalized prior to time \( t \), respectively. Constraint~\eqref{eq:historical_route} ensures that all such historical routing and repair decisions are preserved and treated as fixed inputs in the current optimization.

\subsubsection{Power flow constraints}
The power flow constraints are modeled in (\ref{eq:DN_power}), with component status determined by (\ref{eq:line_state_after_repair}). 

This rolling optimization framework enables the power repair plan to dynamically adapt to evolving gas availability and network topology, while preserving prior routing decisions. By integrating updated constraints and maintaining consistency with past actions, it supports coordinated, multi-stage restoration across interdependent infrastructures. The pseudocode is presented in Algorithm~\ref{alg:rolling_power}.

\begin{algorithm}[htbp]
\caption{Rolling Optimization for Power Crew Scheduling}
\label{alg:rolling_power}
\begin{algorithmic}[1]
\STATE \textbf{Input:} Current time step $t$; prior routing and arrival decisions $\{x_{ij}^{k,\text{prior}}, \tau_j^{k,\text{prior}}\}$; estimated gas supply profile $\hat{W}^{DG}_{m,\tau}$
\STATE Fix all historical decisions for $\tau_j^{k,\text{prior}} \leq t$ by enforcing constraint~\eqref{eq:historical_route}
\STATE Update generator availability constraints using~\eqref{eq:gas_power_conv} based on $\hat{W}^{DG}_{m,\tau}$
\STATE Construct and solve the rolling optimization problem~\eqref{eq:power_crew_rolling_opt}.
\STATE \textbf{Output:} Updated crew schedule, including routing decisions $x_{ij}^k$, arrival times $\tau_j^k$, and repair status indicators $h_{i,t}$ for all $\tau > t$
\end{algorithmic}
\end{algorithm}

\subsection{Approximate flow evaluation for IEGDS} \label{sec:flow_approx}

The proposed BTS algorithm requires fast simulation of the restoration process across a large number of scenarios. However, solving optimal power and gas flows at every simulation step is computationally intensive and impractical for online decision-making. To address this challenge, we introduce a deterministic algebraic approximation method termed Value-Prioritized Flow Allocation (VPFA). This method prioritizes load supply based on network accessibility and load importance, while bypassing nonlinear flow optimization. This enables efficient and scalable estimation of energy loss throughout belief tree simulation. The algorithm proceeds as follows:
\begin{enumerate}
    \item \textbf{Power Island Detection}: Identify power islands based on the status of lines and DGs. For each DG, determine its reachable load set $\mathcal{L}_g$ using breadth-first search.
    
    \item \textbf{Gas Node Valuation}: Each gas node \( j \) is assigned an importance score \( \Phi_j \). For a node with gas load, \( \Phi_j = c^W_j \), the unit gas load value. For a node supplying gas to DGs, \( \Phi_j = \max_i c^P_i / \beta_j \), where \( \max_i c^P_i \) is the highest unit power load among buses \( i \) served by generator \( j \), and \( \beta_j \) is the energy conversion efficiency in (\ref{eq:gas_power_conv}).
    
    \item \textbf{Gas Allocation}: Within each gas island, distribute gas to nodes in descending order of $\Phi_j$ until the available supply is exhausted.
    
    \item \textbf{Power Allocation}: For each DG, compute available power based on gas input, and dispatch it to reachable loads in descending order of $c^P_i$.
    
    \item \textbf{Correction}: Reallocate gas based on actual DG output, and adjust gas allocation to loads accordingly.
\end{enumerate}

A summary of the VPFA procedure is provided in Algorithm~\ref{alg:VPFA}. Although this algebraic approximation is less precise than full optimal flow models, it offers significant computational advantages. Its high speed and ease of parallelization make it particularly well-suited for integration with the belief tree search framework in stochastic repair planning.

\begin{algorithm}[ht]
    \caption{Value-Prioritized Flow Allocation (VPFA)}
    \label{alg:VPFA}
    \begin{algorithmic}[1]
    \STATE Detect power and gas islands
    \FOR{each DG $g$}
        \STATE Identify reachable loads $\mathcal{L}_g$
    \ENDFOR
    \FOR{each gas node $j$}
        \IF{$j$ is a gas-fired DG node}
            \STATE $\Phi_j \leftarrow  \max_i c^P_i / \beta_j$
        \ELSE
            \STATE $\Phi_j \leftarrow c^W_j$
        \ENDIF
    \ENDFOR
    \FOR{each gas island}
        \STATE Sort nodes by $\Phi_j$ and allocate gas accordingly
    \ENDFOR
    \FOR{each DG}
        \STATE Determine available generation capacity
        \STATE Dispatch to $\mathcal{L}_g$ by descending $c^P_i$
    \ENDFOR
    \STATE Reallocate gas based on actual DG output
    \end{algorithmic}
\end{algorithm}

\subsection{Decision-making with multiple gas crews} \label{sec:multi_gas_crew}

While the BTS algorithm detailed in Section \ref{section: tree_search} optimizes the dispatch of a single gas crew under partial observability, its direct application to cases involving multiple gas crews presents significant computational challenges due to the combinatorial explosion of potential action combinations. To this end, we propose a sequential optimization approach.

This method iteratively optimizes the target selection for one gas crew at a time. For the currently selected gas crew, the BTS Algorithm \ref{alg:gas_pomdp} is employed. During the simulation phase of BTS, the future schedules of other gas crews are simulated using the base policy described in Section \ref{section: tree_search} (selecting the pipeline $j^*$ that maximizes $\text{LoadGain}_j / \mathcal{R}_j$). Once the action for the current gas crew is determined, this decision is committed, and the process iterates to the next gas crew.

This sequential optimization strategy effectively mitigates combinatorial complexity, making the problem tractable for real-time applications. By integrating the base policy for other crews during simulation, it implicitly captures the interdependence and cooperative aspects of multi-agent restoration. Being a heuristic approximation, it provides a practical and computationally efficient means to manage multiple gas crews in dynamic and uncertain post-disaster environments, striking a balance between optimality and real-time feasibility.

\subsection{Overall Event-Triggered Restoration Framework} \label{section: overall_framework}

\begin{figure}[!h]
    \centering
    \includegraphics[width=0.8\linewidth]{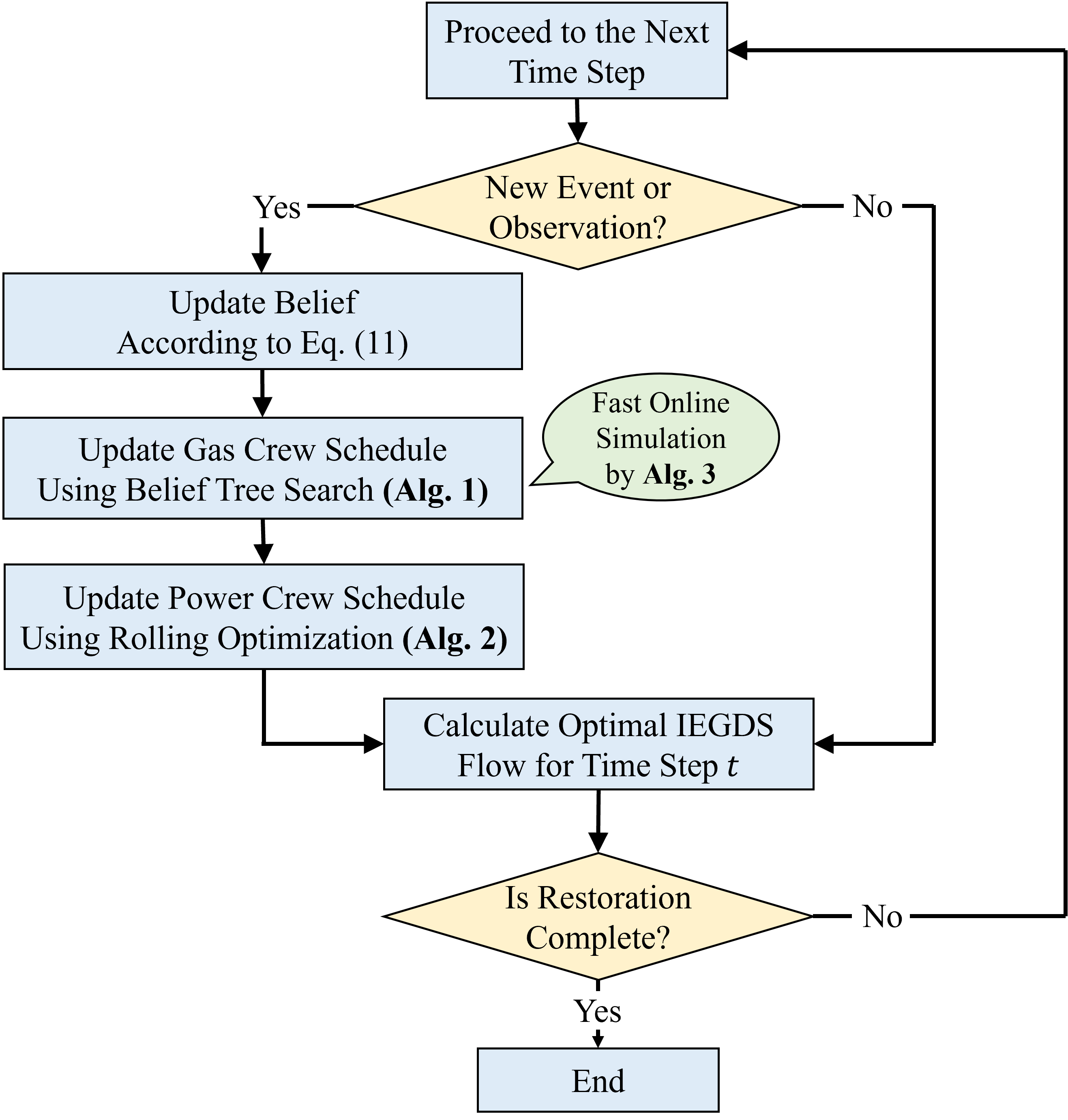}
    \caption{Overall event-triggered restoration framework}
    \label{fig:overall_flowchart}
\end{figure}

To effectively manage the dynamic and uncertain nature of post-disaster restoration, we propose a unified event-triggered decision-making framework. This framework integrates BTS for gas crew scheduling and rolling optimization for power crew dispatch, operating in a continuous, adaptive manner in response to real-time information updates and evolving system states. The overall process is illustrated in Fig. \ref{fig:overall_flowchart}.

At each new time step, if there is a new event (e.g., new fault discovery, fault repair completion, or new observations), the framework initiates a re-optimization cycle. During this cycle, the belief state is updated accordingly. BTS is executed to determine the next actions for gas crews. Meanwhile, the power crew dispatch is re-optimized using Algorithm \ref{alg:rolling_power}, considering the latest network status and gas-fired generator availability. If no event occurs, the system primarily updates the single-period optimal flow calculation. This iterative process continues until all faults are repaired and all unknown pipelines are inspected, leading to a fully restored state.

It is noted that while the simulation phase of the BTS algorithm employs an algebraic approximation of future IEGDS flows using Algorithm 3, the real-time optimal IEGDS flow at the current period $t$ is obtained by solving a single-period instance of problem (\ref{eq:IEGDS_operation}).

It is worth emphasizing that the proposed framework is intrinsically consistent with the operational independence of power and gas utilities. Each utility dispatches its own repair crews, and coordination requires only minimal information exchange at gas-power coupling points (i.e., the availability estimation of gas-fired units and their equivalent demand). As a result, information privacy and dispatch autonomy are rigorously preserved, while effective cross-system coordination is achieved.

\section{Case Study} \label{section: case_study}

To evaluate the performance of the proposed method, we conduct simulation studies on two test systems: a small-scale system with 13 power buses and 7 gas nodes, and a larger system with 123 power buses and 20 gas nodes. The proposed method is implemented using Python 3.11. All experiments are conducted on a laptop equipped with an Intel Core Ultra 5 125H CPU and 16 GB of RAM.

\subsection{Case 1: 13-Power-Node and 7-Gas-Node System} \label{sec:case_13_7}

\subsubsection{Case Description} 
\begin{figure}[!h]
    \centering
    \includegraphics[width=1.0\linewidth]{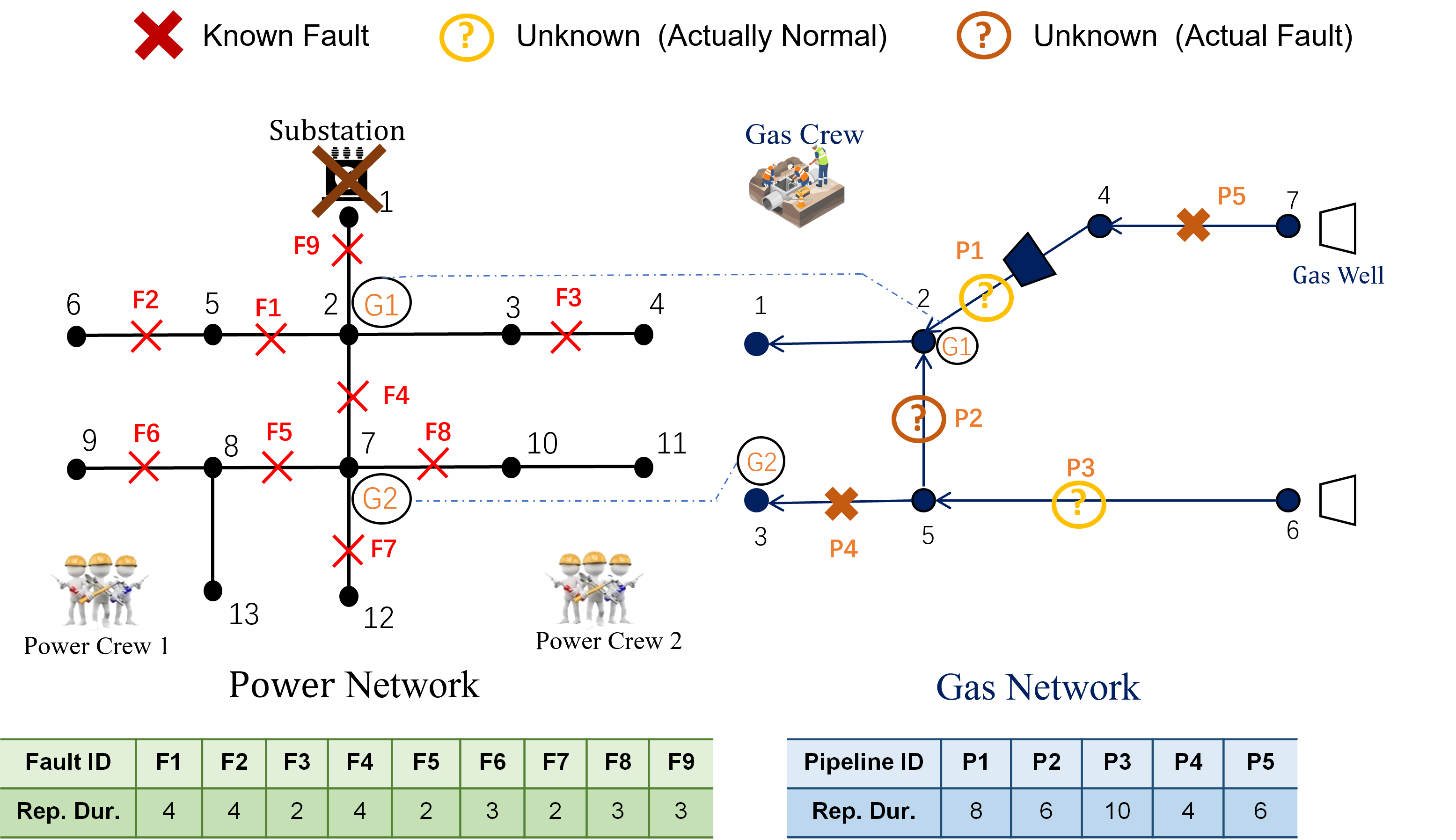}
    \caption{13-power-node and 7-gas-node system }
    \label{fig:case_13_7}
    % \vspace{-5pt}
\end{figure}

As illustrated in Fig.~\ref{fig:case_13_7}, the test system consists of 13 power buses and 7 gas nodes. After an earthquake, line faults F1--F9 are identified in the power network, while the gas network reports two pipelines (P4 and P5) with confirmed faults and three pipelines (P1, P2, and P3) with unknown status. The actual fault condition is that pipeline P2 is damaged, whereas P1 and P3 remain intact. 

Two power crews are assigned to repair the line faults, and one gas crew is responsible for inspection and repair in the gas network. The simulation adopts a time unit of $\Delta t = 0.5$ hour. For gas pipelines with unknown status, each inspection requires one time step.
The repair durations for confirmed faults are shown in Fig.~\ref{fig:case_13_7}, with all values expressed in time steps. For simplicity, the crew travel time between locations is assumed to be proportional to the Euclidean distance. In practical applications, these travel times can be directly obtained from GPS-based navigation systems or real-time traffic data. The complete case data are available in~\cite{data_IEGDS_github}.

\begin{figure}[!h]
    \centering
    \includegraphics[width=0.7\linewidth]{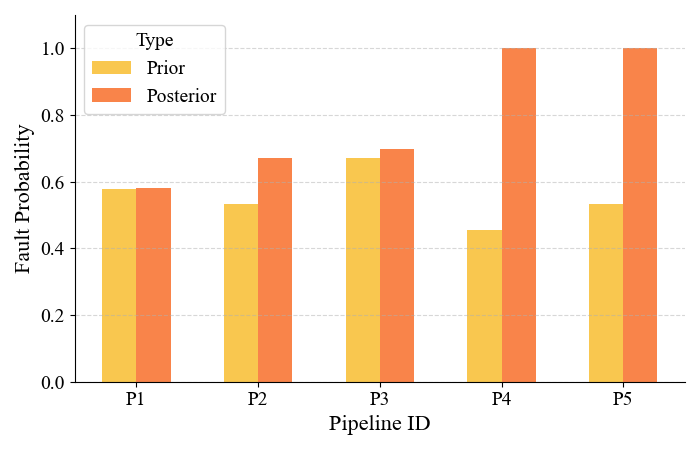}
    \caption{Prior and posterior fault probabilities of pipelines}
    \label{fig:Fault_prob_case1}
    % \vspace{-5pt}
\end{figure}

Fig.~\ref{fig:Fault_prob_case1} shows the pipeline fault probabilities. The prior values are computed using~\eqref{eq:pipe_prior_prob}, and the posterior ones are obtained via~\eqref{eq:pipe_post_prob}. Since pipelines P4 and P5 have been confirmed to be faulty, their posterior probabilities are set to 1. The posterior probabilities can be used for scenario generation.
% Notably, the posterior probability of pipeline P2 increases relative to its prior, as its downstream gas-fired generator G1 experiences a gas shortage, despite the absence of any trouble calls at its upstream gas node 5.

\subsubsection{Performance Comparison of Different Methods} The following approaches are evaluated for gas crew scheduling under partial observability:

\begin{itemize}
    \item \textbf{Proposed Method}:  
    In Algorithm 1, the search depth is set to 2. To model the uncertainty associated with unknown pipeline status, 500 scenarios are generated. Power crews are dispatched by solving problem \eqref{eq:power_crew_rolling_opt}, which is dynamically triggered upon receiving new information from the gas network (e.g., the repair of a pipeline or the restoration of gas supply to a distributed generator).

    \item \textbf{Hindsight Solution}:
    This benchmark assumes full knowledge of all pipeline statuses, which is an ideal case. The power and gas crews are scheduled using deterministic mixed-integer programming \cite{IEGDS_repair_Lin}. Although such perfect information is not attainable in practice, this approach provides a theoretical lower bound on system performance and serves as a reference for evaluating how closely other methods approximate the global optimum.

    \item \textbf{Stochastic Programming (SP)}: A two-stage stochastic programming approach is used for gas crew dispatch, based on 30 scenarios. First-stage decisions determine the crew schedule, while second-stage evaluations assess load restoration under each scenario, conditional on the first-stage decisions. The scenarios are generated using the same probabilities as in the proposed method.

    \item \textbf{Nearest-First Heuristic (NFH)}:  
    The gas crew iteratively selects the nearest unresolved pipeline based on its current location. Power crew scheduling follows the same update procedure as the proposed method.

    \item \textbf{Probability-Based Heuristic (PBH)}:  
    The gas crew prioritizes pipelines with the highest estimated fault probabilities. Power crew scheduling follows the same update procedure as the proposed method.

\end{itemize}

\begin{figure}[!h]
    \centering
    \includegraphics[width=0.8\linewidth]{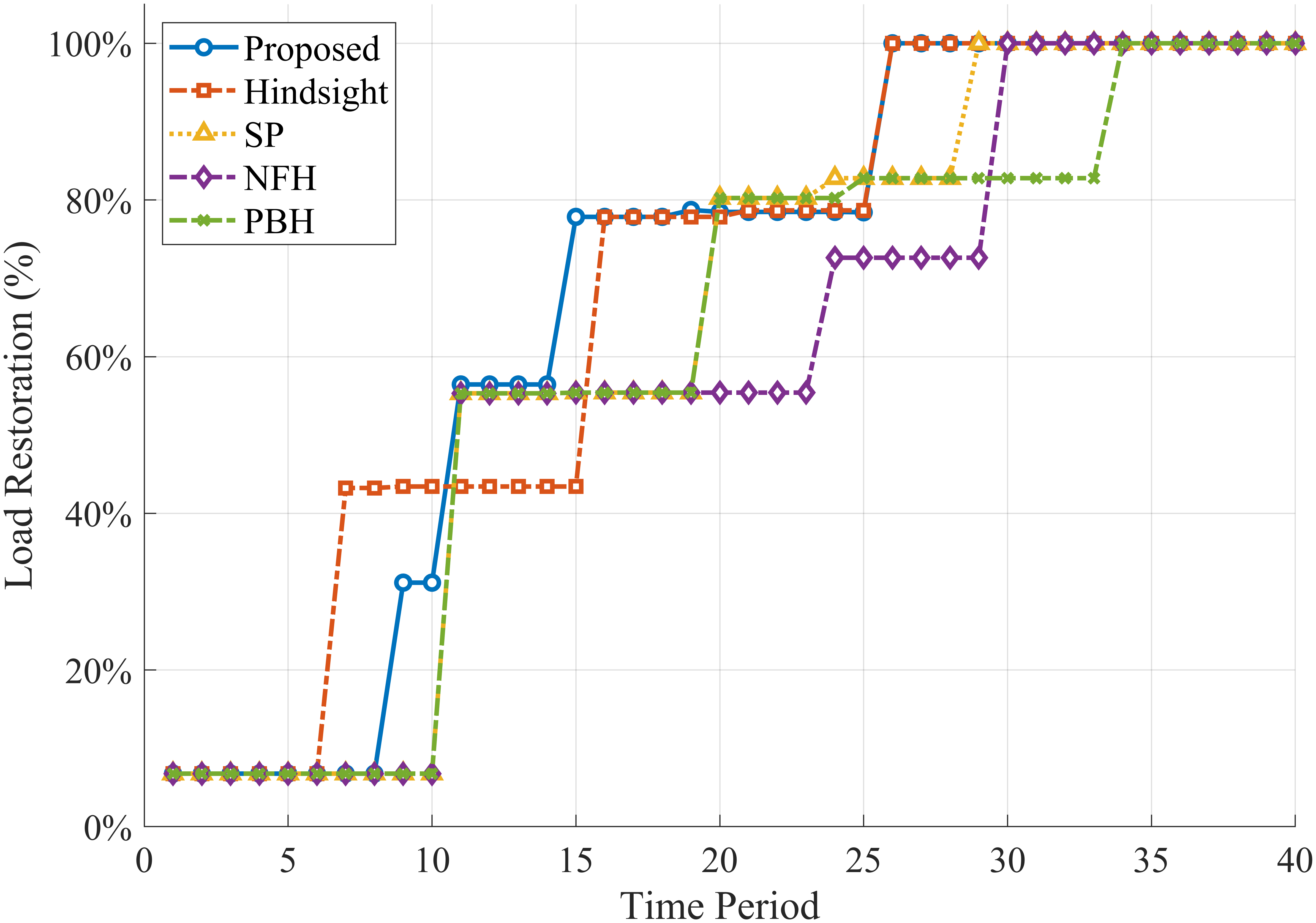}
    \caption{Load restoration curve of each method}
    \label{fig:case_1_restore_curve}
    % \vspace{-5pt}
\end{figure}

Table \ref{tab:method_comparison_case1} summarizes the objective values and the corresponding routing plans for both gas and power crews obtained from different methods. The objective value, Total Outage Cost, represents the cumulative value of shed loads in both the power and gas networks. The relative gap represents the difference in outage cost of each approach compared to the proposed method. The load restoration progress of each method is illustrated in Fig. \ref{fig:case_1_restore_curve}. The primary differences among the methods lie in the routing strategies for the gas crew, as these directly affect the restoration times of gas loads and the resumption of fuel supply to gas-fired units, which in turn influence the power system's load supply.

\begin{table*}[!htbp]
\centering
\scriptsize
\renewcommand{\arraystretch}{1.3}
\caption{Performance Comparison in Case 1}
\begin{tabular}{lcccc}
\hline
\textbf{Method} & \textbf{Total Outage Cost (\$)} & \bd{Relative Gap} & \textbf{Gas Crew Routing} & \textbf{Power Crew Routing} \\
\hline

Proposed Method & 76,602 & - & GC1: P2--P4--P5--P1--P3  &  
\makecell[l]{PC1: F7--F4--F5--F6--F8,\  PC2: F8--F3--F1--F2}  \\

Hindsight Solution & 75,984 & - 0.8\% & \makecell[l]{GC1: P4--P2--P5%--P1--P3  
} &  \makecell[l]{PC1: F7--F5--F4--F1,\ ~~~ PC2: F8--F3--F2--F6--F9} \\

Stochastic Programming (SP) & 88,548 & 15.6\% & GC1: P5--P4--P2--P1--P3  &  \makecell[l]{PC1: F7--F4--F5--F6,\ ~~~ PC2: F3--F8--F1--F2--F9}  \\

Nearest-First Heuristic (NFH) & 99,034 & 29.3\% & GC1: P5--P1--P2--P4--P3  &  \makecell[l]{PC1: F7--F4--F5--F6--F9,\  PC2: F8--F3--F1--F2}  \\

Probability-Based Heuristic (PBH) & 93,783 & 22.4\% & GC1: P5--P4--P3--P2--P1  &  \makecell[l]{PC1: F7--F4--F5--F6,\ ~~~ PC2: F8--F3--F1--F2--F9}  \\

\hline
\end{tabular}
\label{tab:method_comparison_case1}
\end{table*}

As observed from Table \ref{tab:method_comparison_case1}, the proposed method achieves restoration performance comparable to the globally optimal Hindsight Solution, with the main distinction being the order in which pipelines P4 and P2 are accessed. The Hindsight Solution prioritizes repairing the known fault on P4, enabling an earlier recovery of gas unit G2. This leads to a more rapid load restoration during periods 6-9, as shown in Fig. \ref{fig:case_1_restore_curve}. In contrast, the proposed method prioritizes the inspection and repair of pipeline P2, thus re-establishing the gas transmission path of node 6$\rightarrow$5$\rightarrow$2$\rightarrow$1. This strategy enables earlier recovery of gas unit G1 and the critical gas load at node 1, leading to a catch-up and eventual outperformance in restoration progress during periods 11–15 in Fig. \ref{fig:case_1_restore_curve}. Overall, the difference between the proposed method and the Hindsight Solution is marginal, and both strategies are reasonable, demonstrating the effectiveness of the proposed approach.

In comparison, the performance of other benchmark methods is inferior, exhibiting a relative gap of over 15\% compared to the proposed method. The Stochastic Programming method suffers from scalability limitations, making it difficult to account for a large number of possible scenarios within a reasonable computational time. Using only 30 scenarios is insufficient to capture the true joint probability distribution of potential failures, resulting in suboptimal solutions. The Nearest-First Heuristic focuses solely on minimizing travel distance without considering the restoration impact of inspecting or repairing pipelines, leading to poor performance. The Probability-Based Heuristic, while prioritizing pipelines with confirmed faults, fails to account for interdependencies between pipeline failures. As a result, it may overlook critical potential faults, yielding suboptimal outcomes. In contrast, the proposed method effectively leverages scalable online simulations to assess the sequential restoration impact of each pipeline. It employs a tree search-based decision-making process and utilizes a tailored simulation approach that avoids solving complex optimization problems, ensuring both strong optimality and efficiency.

\subsubsection{Detailed Analysis of the Proposed Method} 
\begin{figure}[!h]
	\centering
	\includegraphics[width=1.0\linewidth]{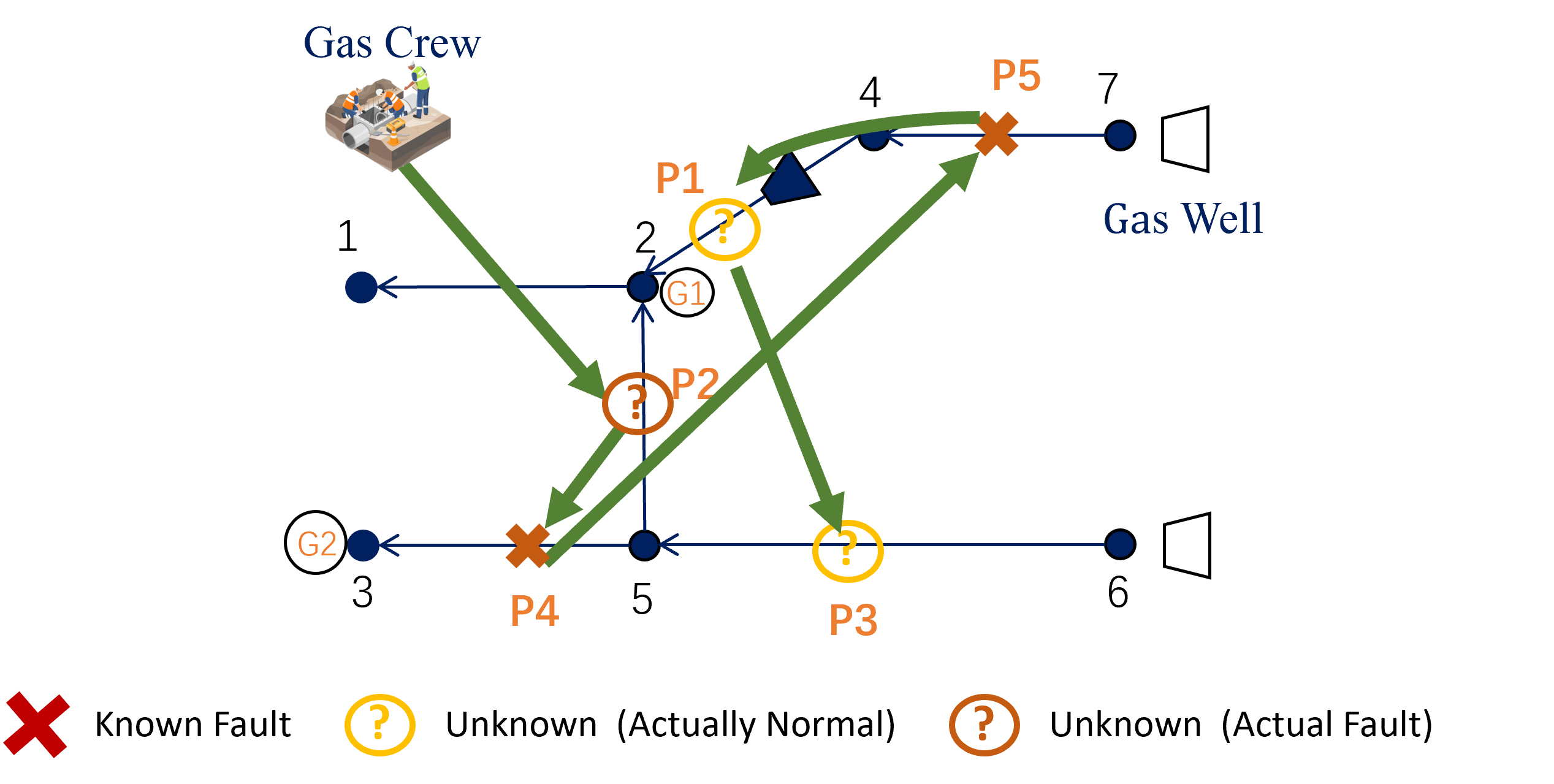}
	\caption{Gas crew routing of the proposed method}
	\label{fig:case_1_major_gas_route}
	% \vspace{-5pt}
\end{figure}

\begin{figure}[!h]
    \centering
    \includegraphics[width=1.0\linewidth]{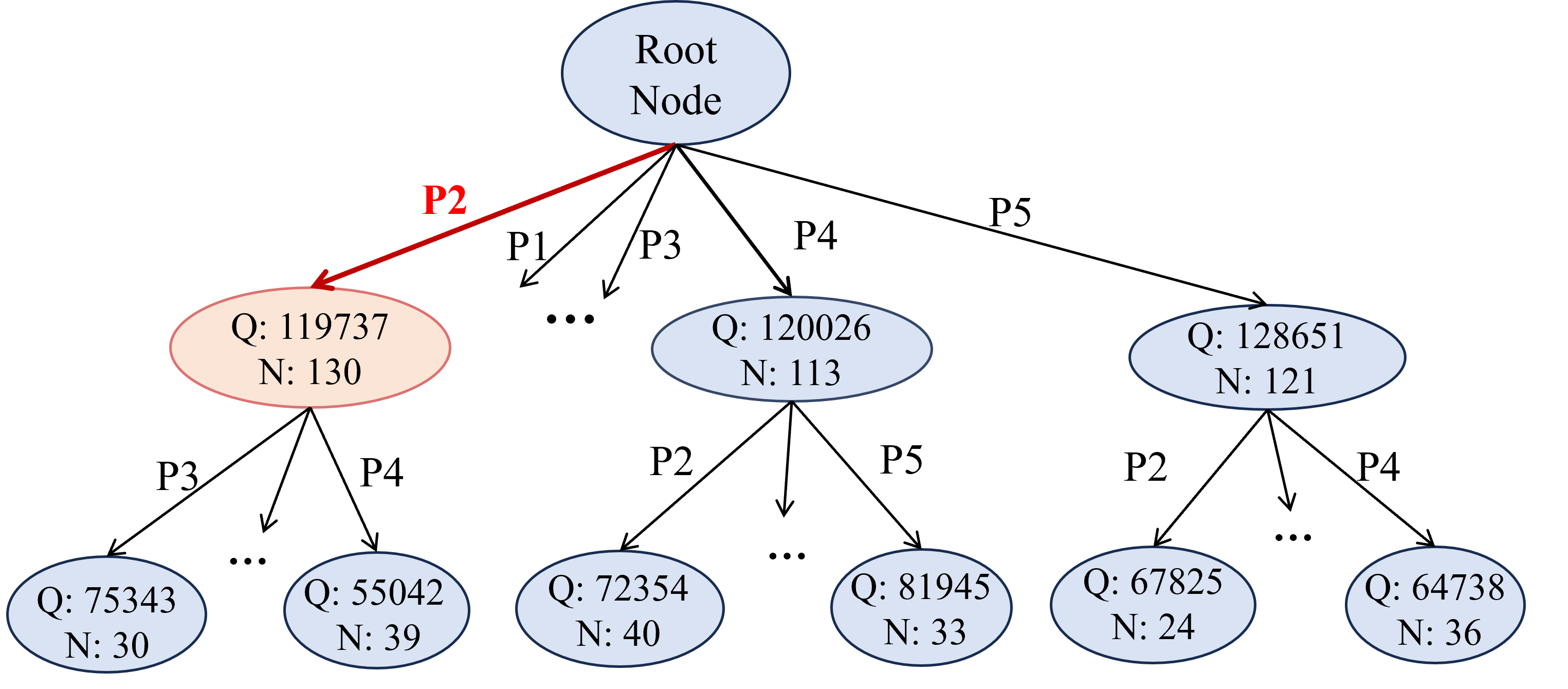}
    \caption{Result of tree search at t=0}
    \label{fig:tree_case1_1}
    % \vspace{-5pt}
\end{figure}

\begin{figure}[!h]
    \centering
    \includegraphics[width=1.0\linewidth]{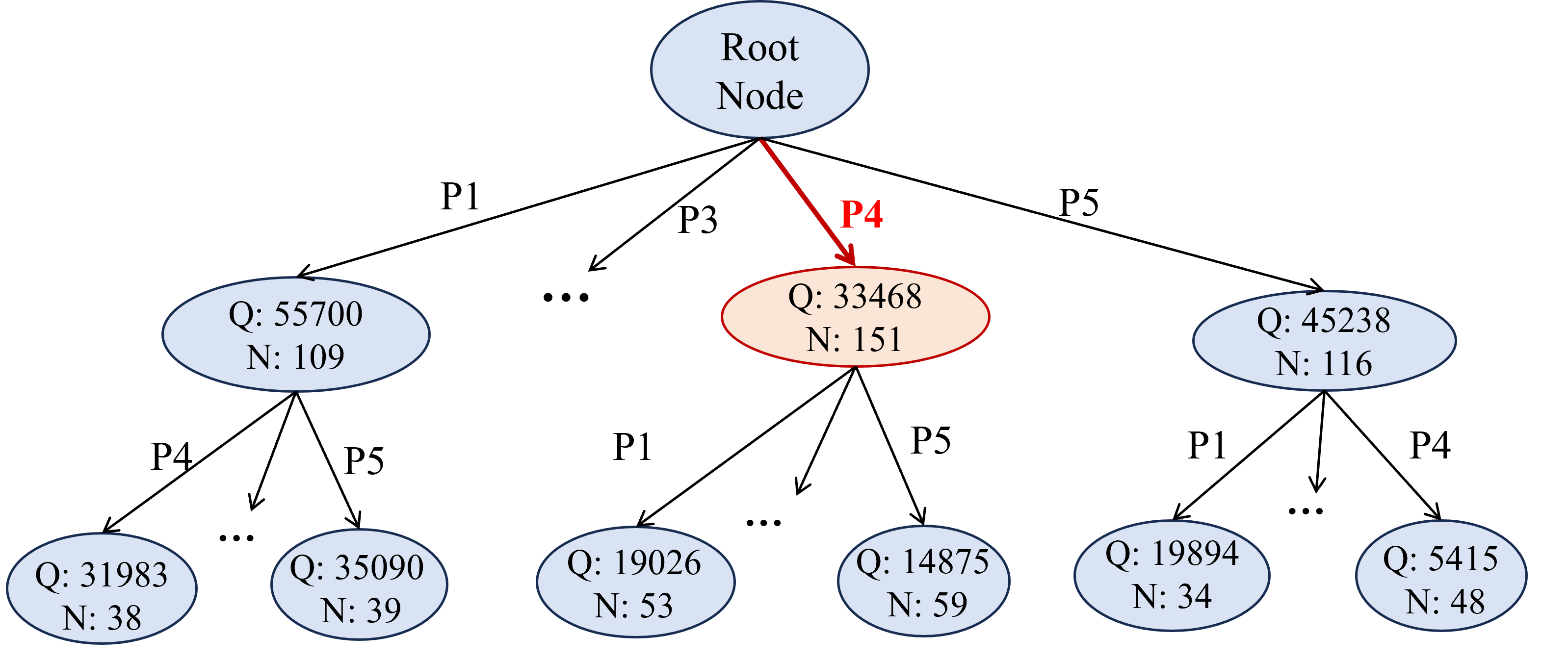}
    \caption{Result of tree search at t=9 after restoring P2}
    \label{fig:tree_case1_2}
    % \vspace{-5pt}
\end{figure}

\begin{figure}[!h]
    \centering
    \includegraphics[width=0.9\linewidth]{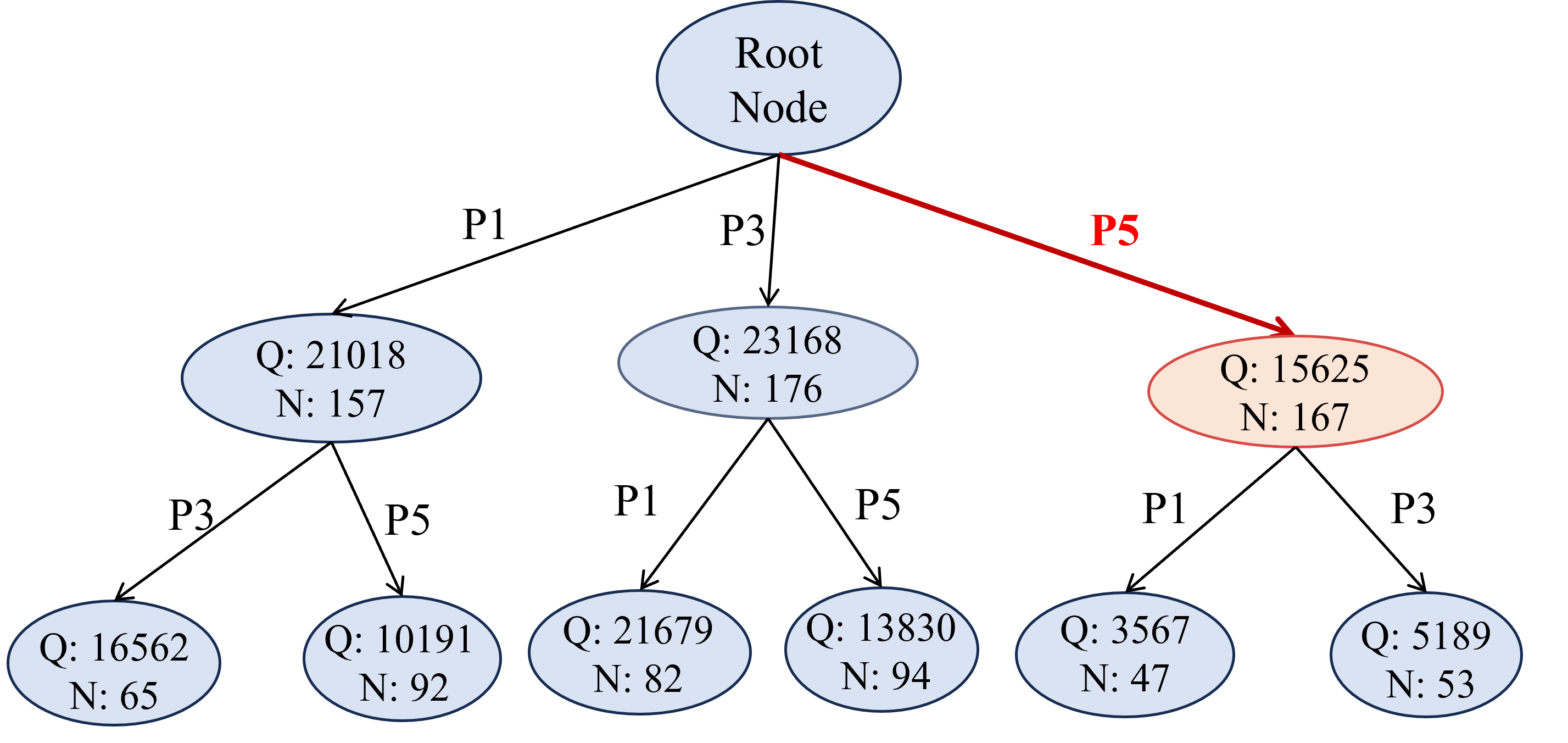}
    \caption{Result of tree search at t=15 after restoring P2 and P4}
    \label{fig:tree_case1_3}
    % \vspace{-5pt}
\end{figure}

To illustrate the decision process of the proposed method, Figs. \ref{fig:tree_case1_1}-\ref{fig:tree_case1_3} present tree search results at three critical decision points, corresponding to the first three targets--P2, P4, and P5--in Fig. \ref{fig:case_1_major_gas_route}. For clarity, only action branches are displayed in the figures; observation branches under different scenarios are omitted. The influence of observations is implicitly captured through simulation results, which are used to estimate the value given a state-action pair, which is called the Q-value in reinforcement learning. For instance, if a pipeline is non-faulty in a sampled scenario, the associated cost only accounts for load shedding during travel and inspection, excluding repair-related losses. To improve readability, only a subset of branches with the lowest Q-values is displayed at each tree level in Figs. \ref{fig:tree_case1_1}-\ref{fig:tree_case1_2}.

Fig. \ref{fig:tree_case1_1} depicts the decision outcome at the initial time step. Each branch in the tree represents the action of accessing a candidate pipeline. At each node, the Q-value quantifies the expected cumulative load loss associated with an action sequence beginning from that decision. The $N$-value specifies the number of simulations used to evaluate the corresponding action.

For the initial decision, selecting pipeline P2 as the first target yields the lowest Q-value (119,737 with 130 simulations), suggesting it is the most promising starting action. At the second level of the tree, following the restoration of P2, pipeline P4 emerges as the most favorable subsequent action, exhibiting the lowest Q-value (55,042) among the remaining candidates. This result is consistent with the actual repair sequence identified by the proposed method and demonstrates the look-ahead capability of the tree search framework.

Fig. \ref{fig:tree_case1_2} illustrates the decision process following the restoration of pipeline P2 by time step 9. At this point, since pipeline P3 is actually intact, the gas transmission path 6$\rightarrow$5$\rightarrow$2 becomes operational, enabling the gas supply to unit G1. As a result, the integrity of P3 can be directly inferred without inspection. Under this condition, the tree search results in Fig. \ref{fig:tree_case1_2} identify P4 as the next optimal target, with the lowest Q-value (33,468 based on 151 simulations), thus supporting the decision to repair P4 at this stage.

In time period 15, as shown in Fig. \ref{fig:tree_case1_3}, P5 is identified as the next optimal target after restoring P2 and P4. Once P5 is repaired, all actual faults in P2, P4, and P5 are resolved, and the gas network is fully restored. Subsequent inspections of P1 and P3 are routine and no longer influence the system's restoration progress.

\subsubsection{Sensitivity Analysis on the Number of Scenarios} 
\begin{figure}[!h]
    \centering
    \includegraphics[width=0.8\linewidth]{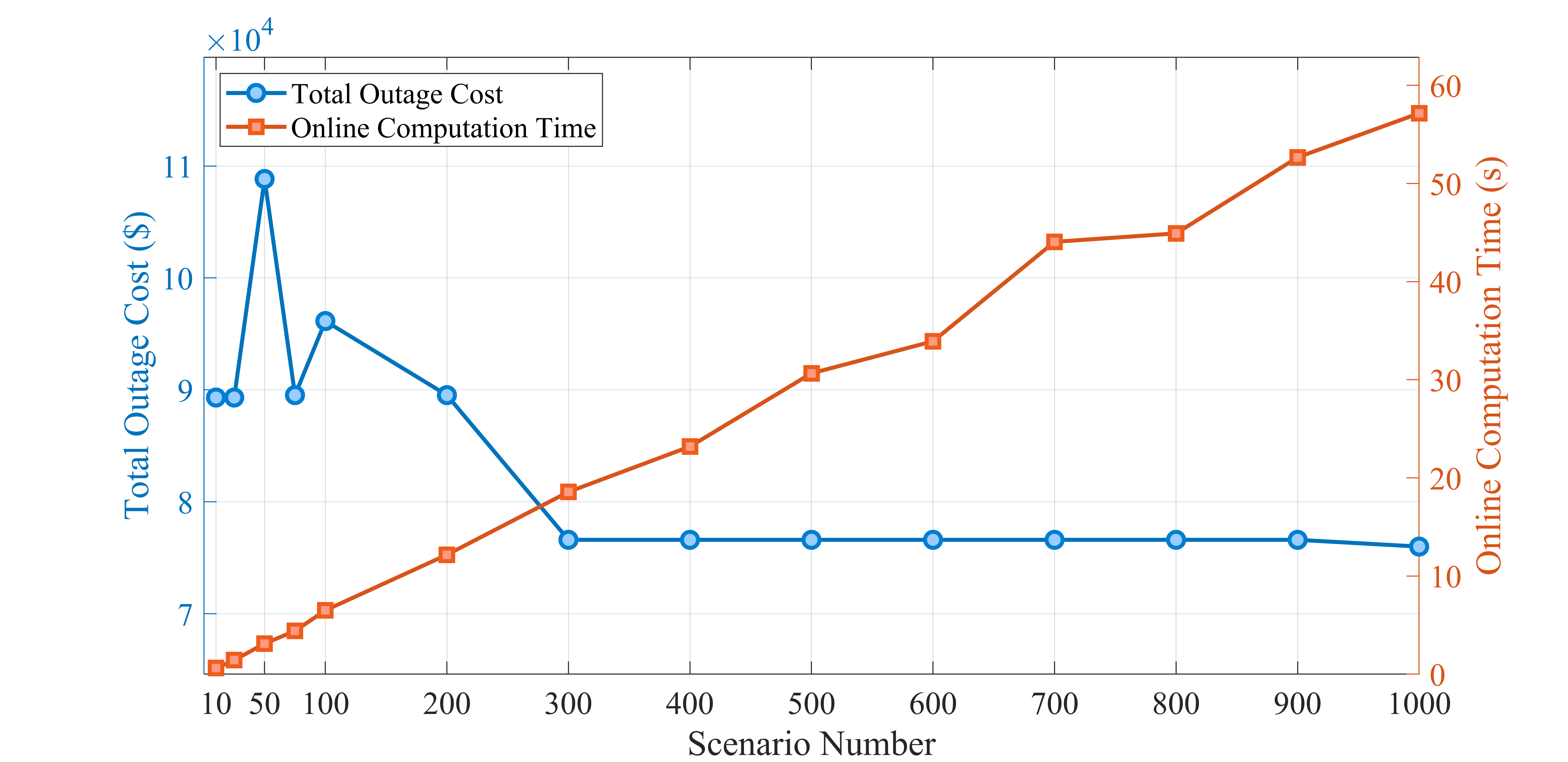}
    \caption{Impact of scenario number on objective and computation time}
    \label{fig:Sensitivity_N_simu_case1_3}
    % \vspace{-5pt}
\end{figure}

The number of simulated scenarios is a critical parameter in the proposed method, as increasing the number of scenarios enhances the accuracy of fault distribution approximation and expected Q-values. Fig. \ref{fig:Sensitivity_N_simu_case1_3} presents the decision performance and online computation time under varying numbers of sampled scenarios. The decision performance is evaluated based on the resulting outage cost, while the online computation time is defined as the maximum duration required to complete a single real-time decision update, i.e., a comprehensive tree search.

As illustrated in Fig. \ref{fig:Sensitivity_N_simu_case1_3}, when the number of scenarios is too small, the decision outcomes are unstable. When the number of scenarios exceeds 300, the objective value stabilizes, indicating that the scenario set is adequate for reliable decision making. The computation time increases approximately linearly with the number of scenarios; nevertheless, even with 1,000 scenarios, the tree search can be completed in 1 minute. These results demonstrate the scalability and computational efficiency of the proposed method.

\subsubsection{Discussion on Computational Complexity}

The proposed BTS algorithm exhibits a distinct computational advantage over conventional stochastic programming approaches. In each decision epoch, the proposed method samples $M_s$ scenarios. For each scenario, a single rollout is simulated, where the tree policy (UCB-based) determines the first $d$ actions and the base policy $\pi_{\text{base}}$ completes the remaining sequence. This design evaluates only a single decision trajectory per scenario rather than exhaustively enumerating all possible action combinations, avoiding exponential growth in complexity. Since both policies rely mainly on simple algebraic operations and the VPFA algorithm introduced in Section~\ref{sec:flow_approx} enables rapid flow simulation, the per-scenario computation is highly efficient. As a result, the total runtime grows approximately linearly with the number of sampled scenarios $M_s$, demonstrating strong scalability.

\begin{table}[h!]
	\centering
	\caption{Comparison of computation time between proposed BTS algorithm and stochastic programming}
	\label{tab:runtime_comparison}
	\begin{tabular}{cc|cc}
		\toprule
		\multicolumn{2}{c|}{\textbf{Proposed BTS algorithm}} & 
		\multicolumn{2}{c}{\textbf{Stochastic programming}} \\
		\midrule
		\textbf{Scenario Count} & \textbf{Time (s)} &
		\textbf{Scenario Count} & \textbf{Time (s)} \\
		\midrule
		200  & 10   & 5   & 315 \\
		400  & 23   & 10  & 848 \\
		600  & 32   & 20  & 2115 \\
		800  & 45   & 30  & 5083 \\
		1000 & 58   & 40  & 8558 \\
		\bottomrule
	\end{tabular}
\end{table}

Table~\ref{tab:runtime_comparison} summarizes the solution time of the proposed BTS algorithm and  stochastic programming under different numbers of sampled scenarios. The BTS method completes each decision update within one minute even under 1,000 scenarios. This efficiency ensures that the framework can perform online rolling updates and dynamically incorporate new post-disaster information in real time.

In contrast, stochastic programming suffers from significant scalability limitations. Its runtime increases exponentially and exceeds 30 minutes when the number of scenarios is greater than 20, rendering it impractical for dynamic updates within the 30-minute decision window. The computational bottleneck stems from the large number of binary routing variables and the replication of network constraints across all scenarios. Moreover, its restriction to a small number of scenarios yields only a coarse approximation of uncertainty, thereby limiting the practical value of the obtained decisions.

Overall, the proposed BTS framework achieves minute-level decision updates, while SP becomes prohibitively slow even under tens of scenarios. These results clearly demonstrate the computational efficiency and real-time applicability of the proposed approach in post-disaster restoration scheduling.

\subsection{Case 2: 123-Power-Node and 20-Gas-Node System}

\subsubsection{Case Description} 
\begin{figure}[!h]
    \centering
    \includegraphics[width=1.0\linewidth]{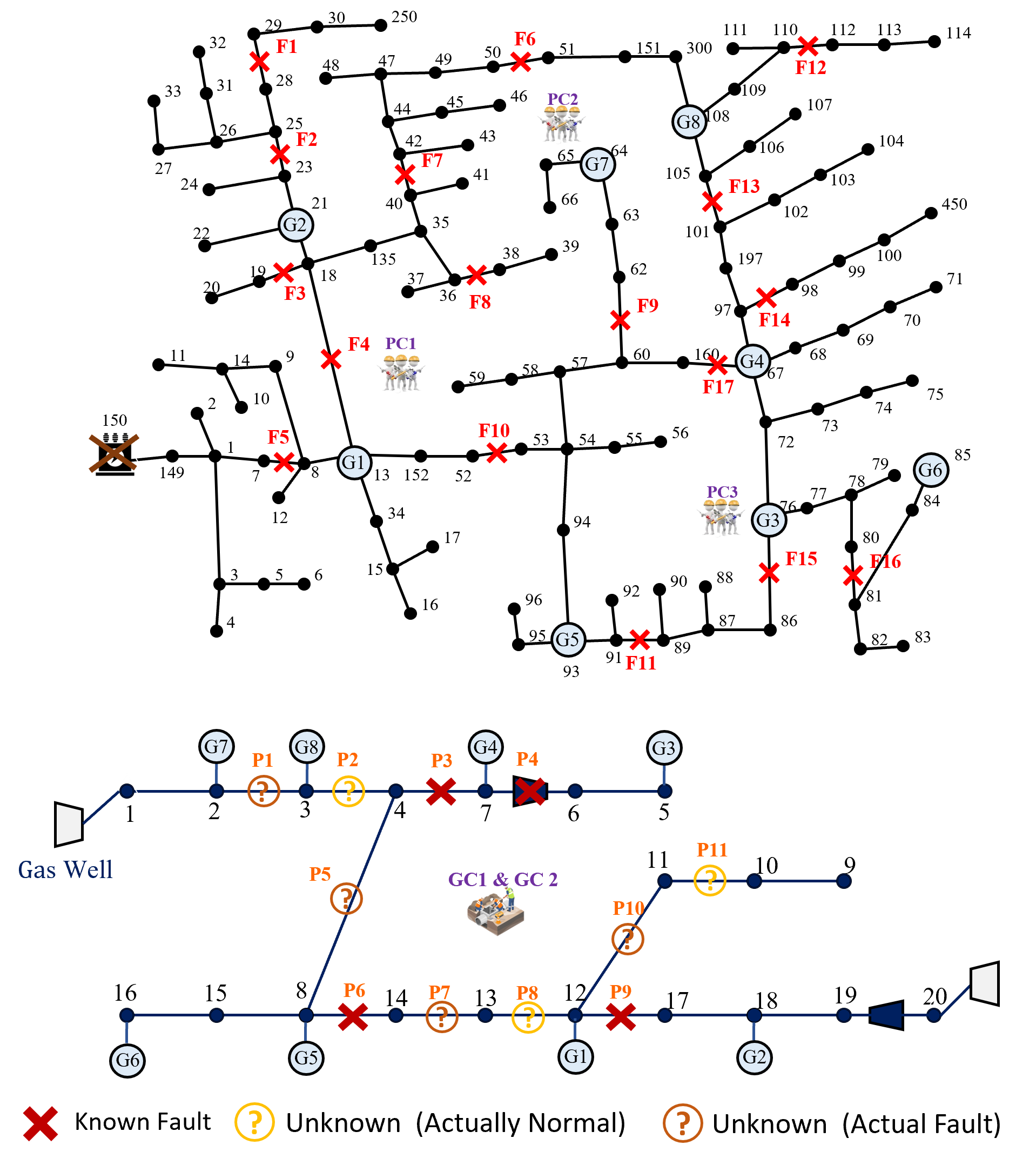}
    \caption{123-power-node and 20-gas-node system }
    \label{fig:case_123_20}
    % \vspace{-5pt}
\end{figure}

Fig.~\ref{fig:case_123_20} presents the test system, which includes 123 power buses, 20 gas nodes, and 8 gas-fired generators. Following a seismic event, the power network experiences 17 identified line faults (F1--F17), while the gas infrastructure reports confirmed failures on pipelines P3, P4, P6, and P9. In addition, the status of seven pipelines--P1, P2, P5, P7, P8, P10, and P11--remains uncertain. The ground truth reveals that among these, P1, P5, P7, and P10 are indeed damaged. Three power crews (PC1, PC2, and PC3) and two gas crews (GC1 and GC2) are assigned for restoration. The simulation operates with a time resolution of $\Delta t = 0.5$ hour. Complete case data can be accessed in~\cite{data_IEGDS_github}.
 % The repair durations for damaged power lines are listed in Table~\ref{tab:power_repair_time_case2}, while those for gas pipelines are detailed in Table~\ref{tab:gas_repair_time_case2}. Each inspection of a pipeline with unknown status takes one time step. 

\subsubsection{Result Analysis} The benchmark methods described in Section~\ref{sec:case_13_7} are applied to this system. Given the large number of pipelines with uncertain status, the proposed method generates 1,000 scenarios to adequately capture possible failure conditions. Due to the system’s scale and complexity, the stochastic programming approach fails to produce a solution within a reasonable time frame and is therefore excluded from this case study.

\begin{figure}[!h]
    \centering
    \includegraphics[width=0.8\linewidth]{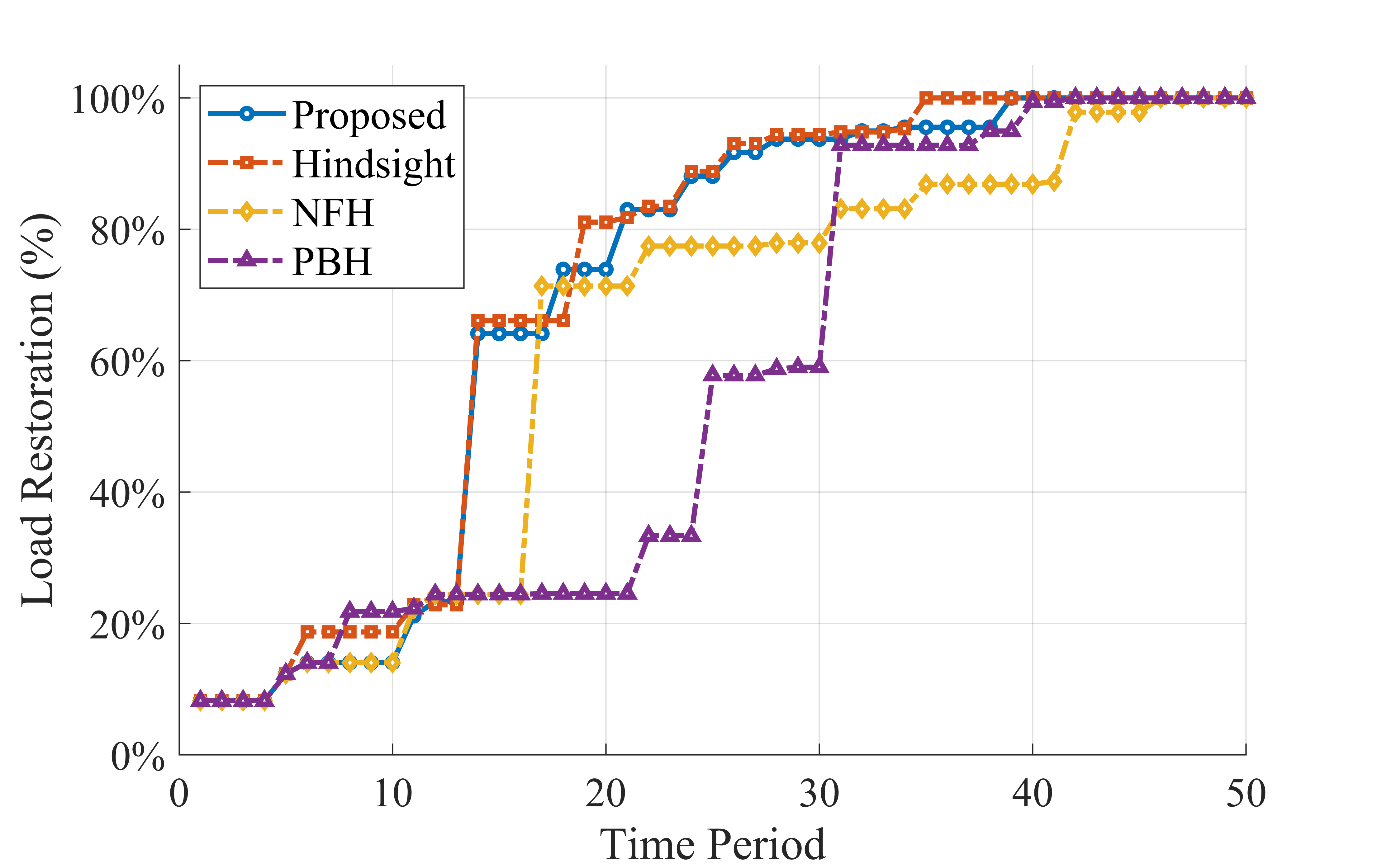}
    \caption{Load restoration curves of case 2}
    \label{fig:case_2_restore_curve}
    % \vspace{-5pt}
\end{figure}

The comparative results are summarized in Table~\ref{tab:method_comparison_case2}, and the corresponding load restoration curves is shown in Fig.~\ref{fig:case_2_restore_curve}. The proposed method achieves a total outage cost of \$61{,}682, which is close to the hindsight optimum of \$59{,}071. In contrast, the Nearest-First Heuristic and Probability-Based Heuristic result in significantly higher costs, with relative gaps exceeding 20\%. These heuristics prioritize travel distance or failure probability alone, while overlooking the downstream effects of pipeline restoration on load recovery.

By explicitly accounting for restoration impacts across all potential scenarios, the proposed method effectively optimizes crew routing and delivers near-optimal performance. As shown in Table~\ref{tab:method_comparison_case2}, the routing plan generated by the proposed method closely resembles the hindsight optimum, with only minor deviations. For instance, after crew GC1 repairs pipeline P1, it cannot infer the status of P2 due to downstream faults in P3 and P5, necessitating additional inspection. A similar situation occurs for GC2 with pipeline P8. Additionally, differences in power crew routing arise from the method's partial observability of the gas network, which prevents precise estimation of generator availability. Despite these uncertainties, the proposed method performs well under limited information.

\begin{table*}[!htbp]
\centering
\scriptsize
\renewcommand{\arraystretch}{2.1}
\caption{Performance Comparison in Case 2}
\begin{tabular}{lcccc}
\hline
\textbf{Method} & \textbf{Total Outage Cost (\$)} & \bd{Relative Gap} & \textbf{Gas Crew Routing} & \textbf{Power Crew Routing} \\
\hline

Proposed Method & 61,682 & - & \makecell[l]{GC1: P1-P2-P3-P4-P7-P6\\ GC2: P5-P9-P8-P10-P11}   &  
\makecell[l]{PC1: F4-F5-F3-F2-F1-F10, ~ PC2: F7-F6-F9-F8-F17-F13  \\ PC3: F11-F15-F16-F14-F12} \\

Hindsight Solution & 59,071 & - 4.2\% & \makecell[l]{GC1: P1-P3-P4-P6\\ GC2: P5-P9-P10-P7}   &  
\makecell[l]{PC1: F4-F5-F2-F1-F3-F13, ~ PC2: F9-F7-F6-F10-F8-F12  \\ PC3: F15-F11-F14-F16-F17}  \\

Nearest-First Heuristic (NFH) & 76,820 & 24.5\% & \makecell[l]{GC1: P7-P6-P5-P2-P1\\ GC2: P8-P9-P10-P11-P4-P3} &  
\makecell[l]{PC1: F4-F5-F2-F1-F3-F10, ~ PC2: F7-F6-F8-F9-F12-F14 \\ PC3: F11-F16-F17-F15-F13} \\

Probability-Based Heuristic (PBH) & 91,462 & 48.3\% & \makecell[l]{GC1: P9-P6-P5-P7-P11\\ GC2: P4-P3-P1-P10-P8-P2} &  
\makecell[l]{PC1: F4-F5-F2-F1-F3-F10, ~ PC2: F7-F6-F8-F9-F12-F13 \\ PC3: F11-F15-F17-F16-F14} \\

\hline
\end{tabular}
\label{tab:method_comparison_case2}
\end{table*}

\begin{figure}[!h]
    \centering
    \includegraphics[width=1.0\linewidth]{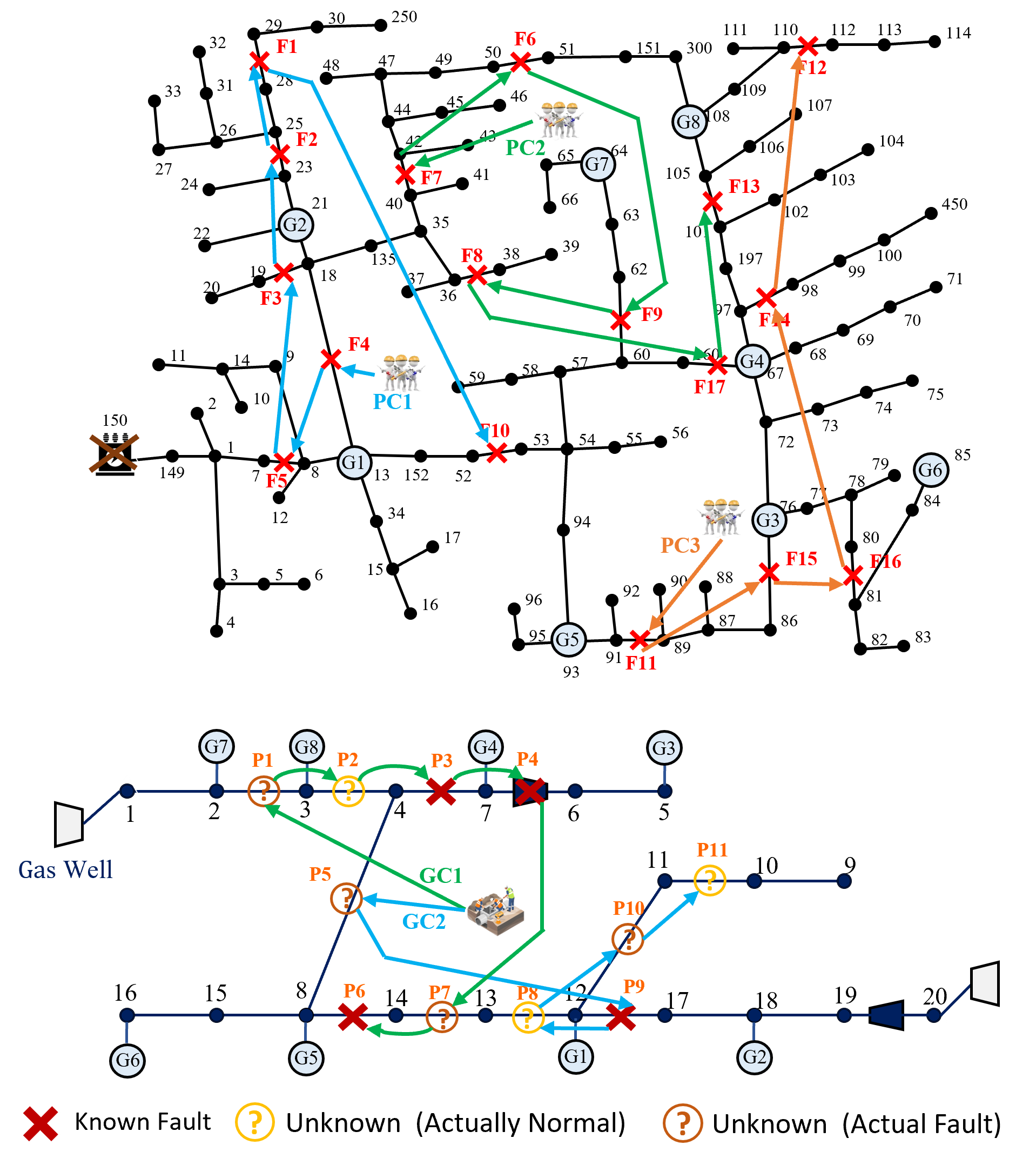}
    \caption{Crew routing of the proposed method in case 2}
    \label{fig:route_123_20}
    % \vspace{-5pt}
\end{figure}

Fig.~\ref{fig:route_123_20} presents the complete routing plan of the proposed method. The initial decisions of the gas crews are especially critical: GC1 inspects and repairs P1, while GC2 targets P5. This enables the restoration of the transmission path of gas node 1$\rightarrow$2$\rightarrow$3$\rightarrow$4$\rightarrow$8$\rightarrow$15$\rightarrow$16, reestablishing gas supply to generators G5, G6, G7, and G8. The corresponding Q-values from the tree search for gas crews GC1 and GC2 are illustrated in Table \ref{tab:tree_case2}. For GC1, inspecting P1 yields the lowest Q-value of 78,361; based on this, GC2 selects P5 with the lowest Q-value of 76,301. The corresponding restoration performance highlights the effectiveness of the proposed method in coordinating multiple gas crews to achieve quick restoration.

\begin{table}[h!]
    \centering
    \scriptsize
    \renewcommand{\arraystretch}{1.3}
    \caption{Initial Tree Search Results for GC1 and GC2}
    \begin{tabular}{c|c|ccccc}
        \toprule
        \multirow{2}{*}{\makecell{\bd{Tree Search} \\ \bd{for GC1}}} & \bd{Action} & \textbf{\textbf{P1}} & P2 & P3 & P6 & P7 \\ \cmidrule(lr){2-7}
        & \bd{Q-value} & \textbf{78,361} & 82,155 & 92,271 & 94,263 & 100,402 \\
        \midrule
        \multirow{2}{*}{\makecell{\bd{Tree Search} \\ \bd{for GC2}}} & \bd{Action} & \textbf{\textbf{P5}} & P8 & P9 & P10 & P11 \\
        \cmidrule(lr){2-7}
        & \bd{Q-value} & \textbf{76,301} & 91,875 & 83,039 & 92,007 & 95,057 \\
        \bottomrule
    \end{tabular}
    \label{tab:tree_case2}
\end{table}

\subsubsection{Sensitivity Analysis of Sequential BTS Optimization} Considering that the sequential BTS introduced in Section III-E optimizes the decisions of multiple gas crews in sequence rather than jointly, it is necessary to examine whether the optimization order affects the overall restoration performance. To this end, three configurations are compared — the original sequential BTS (i.e., optimizing Crew 1 first, followed by Crew 2), a reversed-order BTS (i.e., optimizing Crew 2 first, followed by Crew 1), and the hindsight solution (joint optimization under perfect information)  — to evaluate the sensitivity of the results to the optimization order. The results are presented in Table \ref{tab:sensitivity_order}.

\begin{table}[!htbp]
	\centering
	\scriptsize
	\renewcommand{\arraystretch}{1.3}
	\caption{Sensitivity Analysis of Crew Optimization Order}
	\begin{tabular}{lccc}
		\hline
		\textbf{Method} & \textbf{Objective (\$)} & \textbf{Gap} & \textbf{Gas Crew Routing} \\
		\hline
		Original BTS & 61,682 & – & \makecell[l]{GC1: P1–P2–P3–P4–P7–P6\\ GC2: P5–P9–P8–P10–P11}  \\
		Reversed-order BTS & 62,051 & 0.6\% & \makecell[l]{GC1: P5–P9–P8–P7–P6\\ GC2: P1–P2–P3–P4–P10–P11}  \\
		\makecell[l]{Hindsight Solution \\ (Joint Optimization)} & 59,071 & –4.2\% & \makecell[l]{GC1: P1–P3–P4–P6\\ GC2: P5–P9–P10–P7}   \\
		\hline
	\end{tabular}
	\label{tab:sensitivity_order}
\end{table}

The results show that reversing the optimization order changes the total recovery cost by only 0.6\%, and both sequential configurations remain within 5\% of the hindsight optimum. Given that the hindsight solution benefits from complete state observability unavailable to BTS, the suboptimality attributable solely to the sequential heuristic is even smaller. This negligible difference confirms that the optimization order has only a minor impact on overall performance.

This result can be explained by the fact that, although each crew’s decision is optimized in sequence, the BTS simulation phase inherently accounts for the concurrent actions of other crews following their ongoing tasks or the base policy, thereby preserving inter-crew coordination. Consequently, the sequential optimization structure introduces only a marginal loss of optimality while maintaining computational efficiency.

\subsubsection{Sensitivity Analysis of Scenario Number} 
\begin{figure}[!h]
	\centering
	\includegraphics[width=1.0\linewidth]{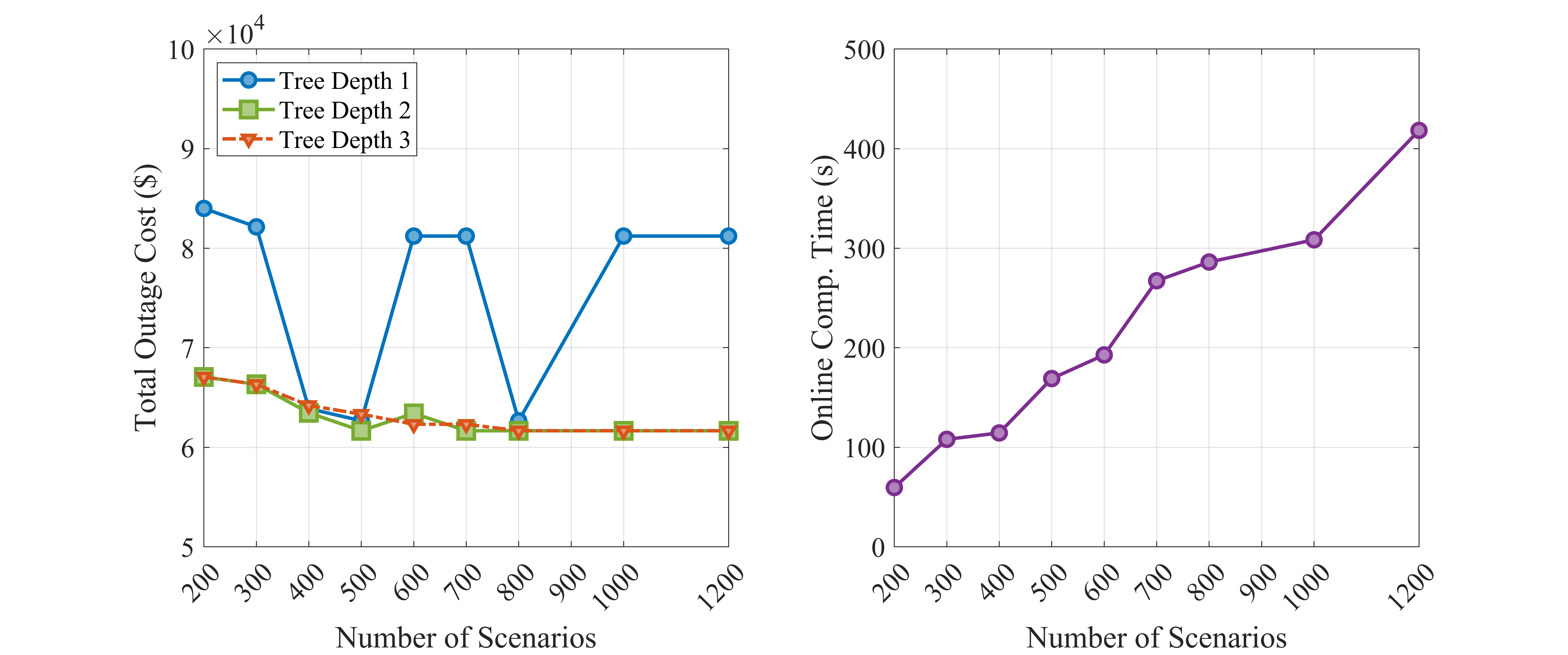}
	\caption{Sensitivity analysis results: (a) Objective value; (b) Online computation time}
	\label{fig:Sensitivity_case2}
	% \vspace{-5pt}
\end{figure}

Compared to Case 1, Case 2 involves a larger set of candidate targets for the gas crew and more complex routing decisions. Consequently, both the number of simulation scenarios and the search tree depth can have a substantial impact on the performance of the proposed method. Fig.~\ref{fig:Sensitivity_case2}(a) presents the objective values under varying tree depths and scenario numbers.

When the tree depth is set to 1, the performance appears to be unstable. In this case, the algorithm performs only one-step optimization, while all subsequent actions are handled by a heuristic policy. This often leads to suboptimal trajectories and inaccurate Q-value estimation due to myopia, particularly in large-scale systems. In contrast, with a tree depth of 2 or 3, the algorithm considers the outcomes of future target selections over multiple steps. This deeper search improves Q-value estimation and leads to more stable performance. Notably, under tree depths of 2 and 3, the results converge when the number of scenarios exceeds 800, demonstrating the robustness of the proposed approach.

Fig.~\ref{fig:Sensitivity_case2}(b) illustrates the online computation time of the proposed algorithm with a tree depth of 2. As shown, even with up to 1,200 scenarios, the computation time remains below 10 minutes. Notably, previous results indicate that 800 scenarios are sufficient for the algorithm to achieve stable and reliable performance, with a corresponding computation time of less than 5 minutes. These results highlight the high efficiency and scalability of the proposed method.

Fig.~\ref{fig:Sensitivity_case2}(b) illustrates the online computation time of the proposed algorithm with a tree depth of 2. As shown, even with up to 1,200 scenarios, the computation time remains below 10 minutes. Notably, previous results indicate that 800 scenarios are sufficient for the algorithm to achieve stable and reliable performance, with a corresponding computation time of less than 5 minutes.  Given the fixed discrete-time step of $\Delta t = 30$ minutes, such computation times are well within the available decision window, indicating that the proposed framework can be readily applied in real post-disaster operations without compromising timeliness or decision quality. These results highlight the high efficiency and scalability of the proposed method.

\begin{comment}
\begin{table}[htbp]
	\centering
	\caption{Online Computation Time under Different Scenario Numbers \red{(TODO)}}
	\label{tab:online_time_case2}
	\begin{tabular}{lcccccc}
		\toprule
		\textbf{Scenario Number} & \textbf{200} & \textbf{400} & \textbf{600} & \textbf{800} & \textbf{1000} & \textbf{1200} \\
		\midrule
		\makecell{\textbf{Online Comp.} \\ \textbf{Time (s)}} & 97.7 & 173.3 & 279.7 & 345.6 & 460.4 & 539.2 \\
		\bottomrule
	\end{tabular}
\end{table}
\end{comment}

\section{Conclusion} \label{section: conclusion}

This paper presents a comprehensive framework for the adaptive restoration of integrated electricity-gas distribution systems under incomplete damage awareness, formulating the problem as a Partially POMDP. To address the challenges of partial observability and real-time decision-making, an innovative BTS algorithm is proposed, enabling gas crews to make informed decisions based on evolving system beliefs. Case studies on two representative IEGDS networks demonstrate the following key advantages of the proposed method:

(1) Decision Quality: The proposed framework achieves restoration performance that closely approaches the ideal hindsight solution and significantly outperforms conventional stochastic programming and heuristic approaches, with a relative gap exceeding 15\%. This highlights its robustness and effectiveness in enhancing the resilience of critical energy infrastructure.

(2) Computational Efficiency: The proposed method demonstrates high computational efficiency, enabling decision updates within 1 minute for Case 1 and within 10 minutes for the large-scale Case 2. This makes it a practical and timely tool for restoration decision-makers, especially during emergency situations with rapidly evolving information.

Future research will explore the integration of data-driven and learning-based techniques to enable offline policy pre-computation, providing pre-trained decision rules that can be promptly applied when communication is temporarily unavailable. Moreover, the proposed framework can be extended toward decentralized or hierarchical coordination, which enables effective restoration planning even under limited communication conditions. These developments will further enhance the practicality, scalability, and resilience of the proposed framework in real post-disaster scenarios. Additionally, future studies will investigate its application to other interdependent infrastructures such as water and communication networks.

\ifCLASSOPTIONcaptionsoff
  \newpage
\fi

\bibliographystyle{IEEEtran}
\bibliography{refs_1, refs_pre-disaster, IEEEabrv}

\end{document}